\pdfoutput=1
\documentclass[12pt,a4paper]{article}

\usepackage{ifthen} 
\newboolean{pdflatex}
\setboolean{pdflatex}{true} 

\newboolean{articletitles}
\setboolean{articletitles}{true} 

\newboolean{uprightparticles}
\setboolean{uprightparticles}{false} 

\def\paperauthors{LHCb collaboration} 
\def\paperasciititle{Search for the lepton-flavour-violating decays B0 to Kstar0 tau e } 
\def\papertitle{Search for the lepton-flavour-violating decays $B^0 \to K^{*0} \tau^\pm e^\mp$} 
\def\paperkeywords{{High Energy Physics}, {LHCb}} 
\def\papercopyright{\the\year\ CERN for the benefit of the LHCb collaboration} 
\def\paperlicence{CC BY 4.0 licence}
\def\paperlicenceurl{https://creativecommons.org/licenses/by/4.0/}

\newif\ifEnableSectionTOCLinks
\EnableSectionTOCLinksfalse 

\usepackage[top=1in, bottom=1.25in, left=1in, right=1in]{geometry}

\usepackage{microtype}
\usepackage{lineno}  
\usepackage{xspace} 
\usepackage{caption} 

\usepackage{graphicx}  
\usepackage{color}
\usepackage{colortbl}
\graphicspath{{./figs/}} 

\usepackage{amsmath} 
\usepackage{amssymb}
\usepackage{amsfonts}
\usepackage{upgreek} 

\newcommand*\patchAmsMathEnvironmentForLineno[1]{%
\expandafter\let\csname old#1\expandafter\endcsname\csname #1\endcsname
\expandafter\let\csname oldend#1\expandafter\endcsname\csname
end#1\endcsname
 \renewenvironment{#1}%
   {\linenomath\csname old#1\endcsname}%
   {\csname oldend#1\endcsname\endlinenomath}%
}
\newcommand*\patchBothAmsMathEnvironmentsForLineno[1]{%
  \patchAmsMathEnvironmentForLineno{#1}%
  \patchAmsMathEnvironmentForLineno{#1*}%
}
\AtBeginDocument{%
\patchBothAmsMathEnvironmentsForLineno{equation}%
\patchBothAmsMathEnvironmentsForLineno{align}%
\patchBothAmsMathEnvironmentsForLineno{flalign}%
\patchBothAmsMathEnvironmentsForLineno{alignat}%
\patchBothAmsMathEnvironmentsForLineno{gather}%
\patchBothAmsMathEnvironmentsForLineno{multline}%
\patchBothAmsMathEnvironmentsForLineno{eqnarray}%
}

\usepackage[pdftex,
            pdfauthor={\paperauthors},
            pdftitle={\paperasciititle},
            pdfkeywords={\paperkeywords}]{hyperref}
\usepackage{hyperxmp}
\hypersetup{
    pdfcopyright={Copyright (C) \papercopyright},
    pdflicenseurl={\paperlicenceurl}
}

\usepackage[colorinlistoftodos,textsize=scriptsize]{todonotes}

\usepackage[bottom,flushmargin,hang,multiple]{footmisc}

\usepackage[all]{hypcap} 

\usepackage{xspace}
\usepackage{upgreek}

\def\lhcb   {\mbox{LHCb}\xspace}

\def\MagUp {\mbox{\em Mag\kern -0.05em Up}\xspace}

\ifthenelse{\boolean{uprightparticles}}%
{

 \def\Pnu         {\ensuremath{\upnu}\xspace}
 
 \def\Ppi         {\ensuremath{\uppi}\xspace}

 \def\Ptau        {\ensuremath{\uptau}\xspace}

 \def\Ppsi        {\ensuremath{\uppsi}\xspace}

 \def\PDelta      {\ensuremath{\Delta}\xspace}
 \def\PXi         {\ensuremath{\Xi}\xspace}
 \def\PLambda     {\ensuremath{\Lambda}\xspace}
 \def\PSigma      {\ensuremath{\Sigma}\xspace}
 \def\POmega      {\ensuremath{\Omega}\xspace}
 \def\PUpsilon    {\ensuremath{\Upsilon}\xspace}
 \let\oldPi\Pi
 \def\PPi         {\ensuremath{\oldPi}\xspace}

 \def\PB      {\ensuremath{\mathrm{B}}\xspace}
 \def\PD      {\ensuremath{\mathrm{D}}\xspace}
 \def\PJ      {\ensuremath{\mathrm{J}}\xspace}
 \def\PK      {\ensuremath{\mathrm{K}}\xspace}
 \def\Pb      {\ensuremath{\mathrm{b}}\xspace}
 \def\Pc      {\ensuremath{\mathrm{c}}\xspace}

 \def\Ps      {\ensuremath{\mathrm{s}}\xspace}

 \def\thebaroffset{0.0em}
}
{

 \def\Pnu         {\ensuremath{\nu}\xspace}
 
 \def\Ppi         {\ensuremath{\pi}\xspace}

 \def\Ptau        {\ensuremath{\tau}\xspace}

 \def\Ppsi        {\ensuremath{\psi}\xspace}
 
 \mathchardef\PDelta="7101
 \mathchardef\PXi="7104
 \mathchardef\PLambda="7103
 \mathchardef\PSigma="7106
 \mathchardef\POmega="710A
 \mathchardef\PUpsilon="7107
 \mathchardef\PPi="7105
 \def\PB      {\ensuremath{B}\xspace}
 \def\PD      {\ensuremath{D}\xspace}
 \def\PJ      {\ensuremath{J}\xspace}
 \def\PK      {\ensuremath{K}\xspace}
 \def\Pb      {\ensuremath{b}\xspace}
 \def\Pc      {\ensuremath{c}\xspace}

 \def\Ps      {\ensuremath{s}\xspace}

 \def\thebaroffset{0.18em}
}
\newcommand{\offsetoverline}[2][\thebaroffset]{\kern #1\overline{\kern -#1 #2}}%

\makeatletter
\ifcase \@ptsize \relax
  \newcommand{\miniscule}{\@setfontsize\miniscule{4}{5}}
\or
  \newcommand{\miniscule}{\@setfontsize\miniscule{5}{6}}
\or
  \newcommand{\miniscule}{\@setfontsize\miniscule{5}{6}}
\fi
\makeatother

\DeclareRobustCommand{\optbar}[1]{\shortstack{{\miniscule (\rule[.5ex]{1.25em}{.18mm})}
  \\ [-.7ex] $#1$}}

\def\taum       {{\ensuremath{\Ptau^-}}\xspace}

\def\neu        {{\ensuremath{\Pnu}}\xspace}
\def\neub       {{\ensuremath{\overline{\Pnu}}}\xspace}

\def\neueb      {{\ensuremath{\neub_e}}\xspace}
\def\neut       {{\ensuremath{\neu_\tau}}\xspace}
\def\neutb      {{\ensuremath{\neub_\tau}}\xspace}

\def\squark    {{\ensuremath{\Ps}}\xspace}

\def\cquark    {{\ensuremath{\Pc}}\xspace}

\def\bquark    {{\ensuremath{\Pb}}\xspace}

\def\pion   {{\ensuremath{\Ppi}}\xspace}
\def\piz    {{\ensuremath{\pion^0}}\xspace}
\def\pip    {{\ensuremath{\pion^+}}\xspace}
\def\pim    {{\ensuremath{\pion^-}}\xspace}

\def\kaon    {{\ensuremath{\PK}}\xspace}

\def\KorKbar {\kern \thebaroffset\optbar{\kern -\thebaroffset \PK}{}\xspace}

\def\Kp      {{\ensuremath{\kaon^+}}\xspace}
\def\Km      {{\ensuremath{\kaon^-}}\xspace}

\def\Kstarz  {{\ensuremath{\kaon^{*0}}}\xspace}

\def\Kstar   {{\ensuremath{\kaon^*}}\xspace}

\def\Dbar    {{\ensuremath{\offsetoverline{\PD}}}\xspace}
\def\D       {{\ensuremath{\PD}}\xspace}

\def\DorDbar {\kern \thebaroffset\optbar{\kern -\thebaroffset \PD}\xspace}
\def\Dz      {{\ensuremath{\D^0}}\xspace}
\def\Dzb     {{\ensuremath{\Dbar{}^0}}\xspace}
\def\Dp      {{\ensuremath{\D^+}}\xspace}
\def\Dm      {{\ensuremath{\D^-}}\xspace}

\def\DpDm    {\ensuremath{\Dp {\kern -0.16em \Dm}}\xspace}
\def\Dstar   {{\ensuremath{\D^*}}\xspace}

\def\Dstarm  {{\ensuremath{\D^{*-}}}\xspace}

\def\Ds      {{\ensuremath{\D^+_\squark}}\xspace}

\def\B       {{\ensuremath{\PB}}\xspace}

\def\BorBbar {\kern \thebaroffset\optbar{\kern -\thebaroffset \PB}\xspace}
\def\Bz      {{\ensuremath{\B^0}}\xspace}

\def\Bd      {{\ensuremath{\B^0}}\xspace}

\def\BdorBdbar {\kern \thebaroffset\optbar{\kern -\thebaroffset \Bd}\xspace}

\def\Bs      {{\ensuremath{\B^0_\squark}}\xspace}

\def\BsorBsbar {\kern \thebaroffset\optbar{\kern -\thebaroffset \Bs}\xspace}

\def\jpsi     {{\ensuremath{{\PJ\mskip -3mu/\mskip -2mu\Ppsi}}}\xspace}

\def\Y#1S{\ensuremath{\PUpsilon{(#1S)}}\xspace}

\def\LorLbar     {\kern \thebaroffset\optbar{\kern -\thebaroffset \PLambda}\xspace}

\def\BF         {{\ensuremath{\mathcal{B}}}\xspace}
\def\BR         {\BF}

\newcommand{\decay}[2]{\mbox{\ensuremath{#1\!\to #2}}\xspace}

\def\to                 {\ensuremath{\rightarrow}\xspace}

\def\AT#1     {\ensuremath{A_{\mathrm{T}}^{#1}}\xspace}           

\def\C#1      {\ensuremath{\mathcal{C}_{#1}}\xspace}                       
\def\Cp#1     {\ensuremath{\mathcal{C}_{#1}^{'}}\xspace}                    
\def\Ceff#1   {\ensuremath{\mathcal{C}_{#1}^{\mathrm{(eff)}}}\xspace}        
\def\Cpeff#1  {\ensuremath{\mathcal{C}_{#1}^{'\mathrm{(eff)}}}\xspace}       
\def\Ope#1    {\ensuremath{\mathcal{O}_{#1}}\xspace}                       
\def\Opep#1   {\ensuremath{\mathcal{O}_{#1}^{'}}\xspace}                    

\newcommand{\nospaceunit}[1]{\ensuremath{\text{#1}}}
\newcommand{\aunit}[1]{\ensuremath{\text{\,#1}}}

\newcommand{\tev}{\aunit{Te\kern -0.1em V}\xspace}
\newcommand{\gev}{\aunit{Ge\kern -0.1em V}\xspace}
\newcommand{\mev}{\aunit{Me\kern -0.1em V}\xspace}
\newcommand{\kev}{\aunit{ke\kern -0.1em V}\xspace}
\newcommand{\ev}{\aunit{e\kern -0.1em V}\xspace}

\newcommand{\mevc}{\ensuremath{\aunit{Me\kern -0.1em V\!/}c}\xspace}
\newcommand{\gevc}{\ensuremath{\aunit{Ge\kern -0.1em V\!/}c}\xspace}
\newcommand{\mevcc}{\ensuremath{\aunit{Me\kern -0.1em V\!/}c^2}\xspace}
\newcommand{\gevcc}{\ensuremath{\aunit{Ge\kern -0.1em V\!/}c^2}\xspace}

\def\mm   {\aunit{mm}\xspace}

\def\mum  {\ensuremath{\,\upmu\nospaceunit{m}}\xspace}

\def\fb   {\ensuremath{\aunit{fb}}\xspace}
\def\invfb   {\ensuremath{\fb^{-1}}\xspace}

\def\gsim{{~\raise.15em\hbox{$>$}\kern-.85em
          \lower.35em\hbox{$\sim$}~}\xspace}
\def\lsim{{~\raise.15em\hbox{$<$}\kern-.85em
          \lower.35em\hbox{$\sim$}~}\xspace}

\def\sPlot{\mbox{\em sPlot}\xspace}

\def\pt         {\ensuremath{p_{\mathrm{T}}}\xspace}

\def\ptot       {\ensuremath{p}\xspace}
\def\et         {\ensuremath{E_{\mathrm{T}}}\xspace}

\def\evtgen     {\mbox{\textsc{EvtGen}}\xspace}

\def\geant      {\mbox{\textsc{Geant4}}\xspace}

\def\photos     {\mbox{\textsc{Photos}}\xspace}

\def\pythia     {\mbox{\textsc{Pythia}}\xspace}

\def\tell1  {TELL1\xspace}
\def\ukl1   {UKL1\xspace}

\newcommand{\ie}{\mbox{\itshape i.e.}\xspace}

\newcommand{\lhcborcid}[1]{\href{https://orcid.org/#1}{\hspace*{0.1em}\raisebox{-0.45ex}{\includegraphics[width=1em]{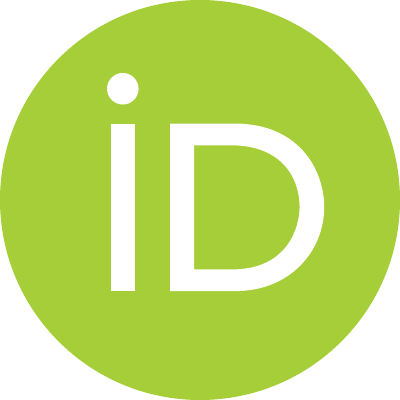}}}}

\hypersetup{
  colorlinks   = true, 
  urlcolor     = blue, 
  linkcolor    = blue, 
  citecolor    = red   
}

\ifEnableSectionTOCLinks
    \usepackage[explicit]{titlesec} 
    
    \let\oldcontentsline\contentsline
    \renewcommand

    \titleformat{\section}{\normalfont\Large\bf}{\hyperlink{tocsection.\thesection}{{\thesection} \parbox[t]{\dimexpr\textwidth-1pc}{#1}}}{1pc}{}

    \titleformat{\subsection}{\normalfont\bf}{\hyperlink{tocsubsection.\thesubsection}{{\thesubsection} \parbox[t]{\dimexpr\textwidth-1pc}{#1}}}{1pc}{}

    \titleformat{name=\section,numberless}[display]{}{}{0pt}{\normalfont\Huge\bfseries #1}
\fi

\usepackage{cite} 
\usepackage{mciteplus}

\def\tauola     {\mbox{\textsc{Tauola}}\xspace}

\mathchardef\mhyphen="2D
\newcommand{\kfolding}{\ensuremath{k\mhyphen\mathrm{folding}\xspace}}
\newcommand{\SC}{\mymath{\decay{\Bd}{\Kstarz\tau^-e^+}}}
\newcommand{\OC}{\mymath{\decay{\Bd}{\Kstarz\tau^+e^-}}}
\newcommand{\SCOC}{\mymath{\decay{\Bd}{\Kstarz\tau^{\mp}e^{\pm}}}}

\renewcommand{\epsilon}{\varepsilon}
\renewcommand{\theta}{\vartheta}

\newcommand{\mymath}[1]{\ensuremath{#1}\xspace}

\newcommand{\DToKPiPi}{\mymath{\decay{\Dm}{\Kp\pim\pim}}}

\newcommand{\DspToKKPi}{\mymath{\decay{\Ds}{\Kp\Km\pip}}}

\newcommand{\osTauMu}{\mymath{\tau^\pm\mu^\mp}}

\newcommand{\KstToKPi}{\mymath{\decay{\Kstarz}{\Kp\pim}}}

\newcommand{\osBToKstTauM}{\mymath{\decay{\Bd}{\Kstarz\osTauMu}}}

\newcommand{\tauToPiPiPi}{\mymath{\decay{\taum}{\pim\pip\pim\neut}}}
\newcommand{\tauToPiPiPiPiz}{\mymath{\decay{\taum}{\pim\pip\pim\piz\neut}}}

\newcommand{\BdToDDs}{\mymath{\decay{\Bd}{\Dm\Ds}}}

\usepackage{enumitem}
\usepackage{cleveref}
\usepackage{multirow}
\usepackage{placeins}
\usepackage{tikz}
\usetikzlibrary{trees,arrows,automata,shapes}
\usetikzlibrary{decorations.pathmorphing}
\usetikzlibrary{decorations.markings}
\usetikzlibrary{shapes.symbols}
\usetikzlibrary{shapes.geometric}
\usetikzlibrary{shapes.arrows}
\usetikzlibrary{positioning}
\tikzset
{
   vector/.style = {decorate, decoration={snake,amplitude=2pt, segment length=5pt}},
   fermion/.style = {postaction={decorate}, decoration={markings,mark=at position .55 with {\arrow{>}}}},
   fermionbar/.style = {postaction={decorate}, decoration={markings,mark=at position .55 with {\arrow{<}}}},
   gluon/.style = {decorate, decoration={coil,amplitude=3pt, segment length=5pt}},
   scalar/.style = {dashed, postaction={decorate}, decoration={markings,mark=at position .55 with {\arrow{>}}}},
   scalarbar/.style = {dashed, postaction={decorate}, decoration={markings,mark=at position .55 with {\arrow{<}}}}
}
\tikzstyle{information text}=[draw,rounded corners,inner sep=1ex]
\usepackage{longtable}

\usepackage{longtable} 

\begin{document}

\renewcommand{\thefootnote}{\fnsymbol{footnote}}
\setcounter{footnote}{1}


\begin{titlepage}
\pagenumbering{roman}

\vspace*{-1.5cm}
\centerline{\large EUROPEAN ORGANIZATION FOR NUCLEAR RESEARCH (CERN)}
\vspace*{1.5cm}
\noindent
\begin{tabular*}{\linewidth}{lc@{\extracolsep{\fill}}r@{\extracolsep{0pt}}}
\ifthenelse{\boolean{pdflatex}}
{\vspace*{-1.5cm}\mbox{\!\!\!\includegraphics[width=.14\textwidth]{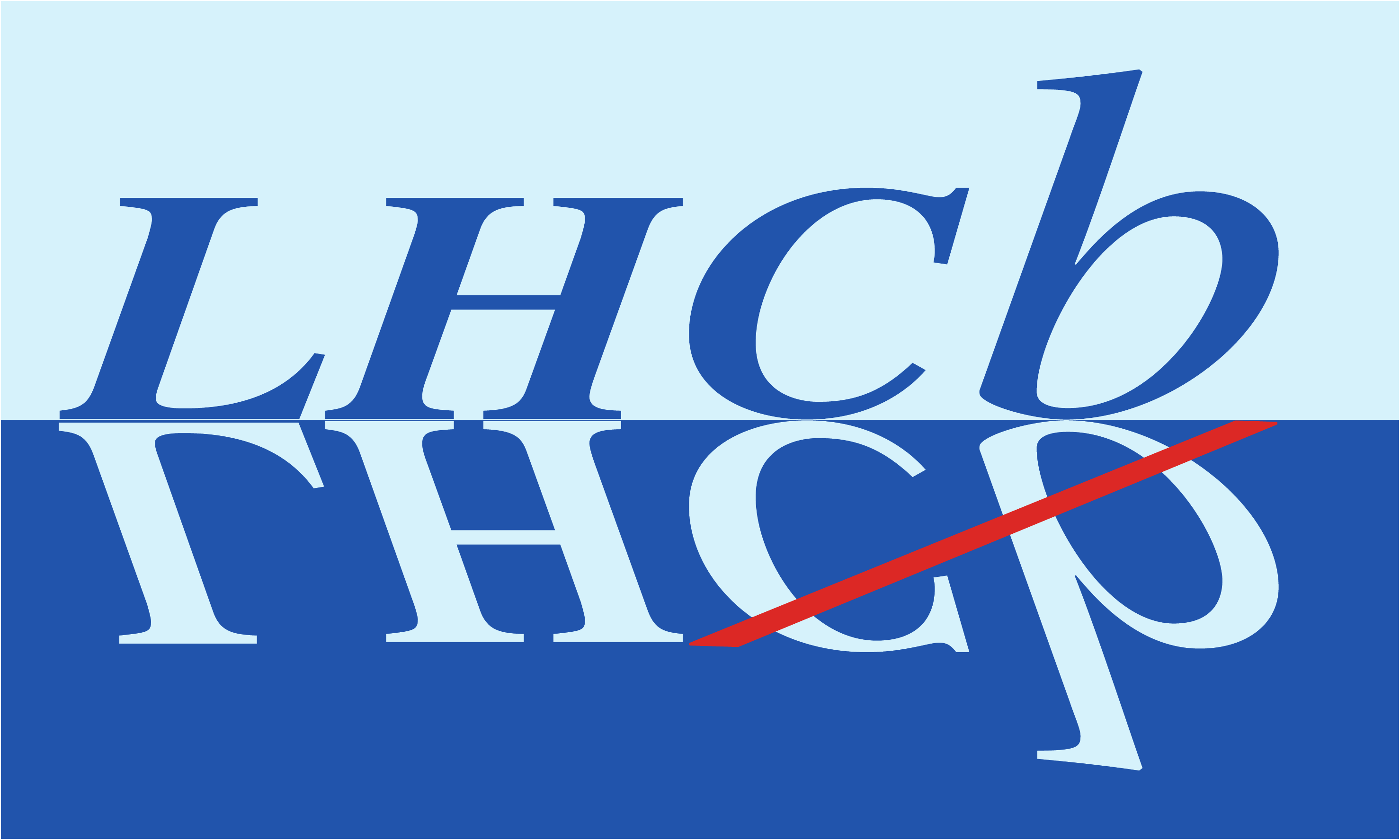}} & &}%
{\vspace*{-1.2cm}\mbox{\!\!\!\includegraphics[width=.12\textwidth]{figs/lhcb-logo.pdf}} & &}%
\\
 & & CERN-EP-2025-107 \\  
 & & LHCb-PAPER-2025-005 \\  
 & & December 5, 2025 \\
 & & \\
\end{tabular*}

\vspace*{4.0cm}

{\normalfont\bfseries\boldmath\huge
\begin{center}
  \papertitle 
\end{center}
}

\vspace*{2.0cm}

\begin{center}
\paperauthors\footnote{Authors are listed at the end of this paper.}
\end{center}

\vspace{\fill}

\begin{abstract}
  \noindent
  A first search at LHCb for the lepton-flavour-violating decays $B^0\to K^{*0}\tau^\pm e^\mp$ is presented.  The analysis is performed using a sample of proton-proton collision data, collected with the LHCb detector at a centre-of-mass energy of 13 TeV between 2016 and 2018,  corresponding to an integrated luminosity of 5.4 fb$^{-1}$. No significant signal is observed, and upper limits on the branching fractions are determined to be \mbox{$\cal{B}$$(B^0 \to K^{*0}\tau^-e^+)< 5.9$ $(7.1)\times 10^{-6}$} and \mbox{$\cal{B}$$(B^0 \to K^{*0}\tau^+e^-)< 4.9$ $(5.9)\times 10^{-6}$} at the 90\% (95\%) confidence level. These results correspond to the current most stringent upper limits for $b \to s\tau l$ transitions.

\end{abstract}

\vspace*{2.0cm}

\begin{center}
Published in JHEP 11 (2025) 172
(26 November 2025,
    Volume 2025, article number 172)
\end{center}

\vspace{\fill}

{\footnotesize 
\centerline{\copyright~\papercopyright. \href{\paperlicenceurl}{\paperlicence}.}}
\vspace*{2mm}

\end{titlepage}


\newpage
\setcounter{page}{2}
\mbox{~}
%
%
%
%


\renewcommand{\thefootnote}{\arabic{footnote}}
\setcounter{footnote}{0}


\cleardoublepage


\pagestyle{plain} 
\setcounter{page}{1}
\pagenumbering{arabic}


\section{Introduction
\label{sec:Introduction}}
Lepton-flavour violation (LFV) has been observed in neutral leptons through neutrino oscillations~\cite{Super-Kamiokande:1998kpq, SNO:2001kpb, OPERA:2010pne, T2K:2013bqz, NOvA:2016vij, DoubleChooz:2011ymz, DayaBay:2012fng, RENO:2012mkc}. For charged leptons, however, the Standard Model~(SM) predicts that   LFV is 
 beyond any current experimental sensitivity~\cite{Raidal:2008jk}. 
Thus, any observation of a charged LFV decay would provide clear evidence for physics beyond the SM.    

The LHCb collaboration has performed searches for lepton-flavour-violating $b$-hadron decays into final states with an electron and a muon 
~\cite{LHCb-PAPER-2017-031,LHCb-PAPER-2019-022, LHCb-PAPER-2022-008}, 
and with a tau and a muon 
~\cite{LHCb-PAPER-2019-016, LHCb-PAPER-2019-043, LHCb-PAPER-2022-021, LHCb-PAPER-2024-006}. No searches at LHCb have been conducted involving a tau and an electron.  Analogous searches have been  performed by other experiments, finding no signal~\cite{  BaBar:2006tnv, 
BaBar:2007xeb,
BELLE:2011bej,
BaBar:2012azg,
BaBar:2013swg,
BELLE:2019xld,
Belle:2021rod,Belle:2022pcr,  Belle:2023jwr}. 
The most stringent limit on  the rate of a $b \to  s \tau e$ transition  is  \mbox{$\BR(B^+ \to K^+ \tau^\pm e^\mp) < 1.5 \times 10^{-5}$} at the 90\% confidence level, set by the Belle collaboration~\cite{Belle:2022pcr}. For the decay
\mbox{$\Bz \to \Kstar(892)^0 \tau^\pm e^\mp$}, an upper limit is set by the Belle collaboration to \mbox{$\cal{B}$$(B^0 \to K^{*0}\tau^-e^+)< 4.4$ $\times 10^{-5}$} and \mbox{$\cal{B}$$(B^0 \to K^{*0}\tau^+e^-)< 2.8$ $\times 10^{-5}$} at the 90\% confidence level~\cite{Belle:2025sgv}.
In this article, the search for the charged lepton-flavour-violating decay \mbox{$\Bz \to \Kstar(892)^0 \tau^\pm e^\mp$} at LHCb is presented. The inclusion of charge-conjugate processes is implied throughout.   
The \mbox{$\Kstar(892)^0$} (referred to in the text as \Kstarz) meson is reconstructed through its decay into a kaon and a pion. 
The decays \OC with the charged kaon and the electron having opposite charges and \SC with the charged kaon and the electron having the same charge are treated independently. These channels could be affected differently by extensions beyond the SM ~\cite{Bordone:2019uzc}, and they are contaminated by different background contributions. 
The $\tau$ lepton is reconstructed through the decays \tauToPiPiPi or \tauToPiPiPiPiz for which the \piz meson is not explicitly reconstructed, representing a combined branching ratio of approximately 14\%. 
The decay \BdToDDs followed by  \DToKPiPi and \DspToKKPi, 
sharing six tracks in the final states and with a precisely known branching fraction, is used both as a normalisation and control channel to test the reliability of the simulation and to evaluate the relevant systematic uncertainties. 
 
This analysis is performed on a dataset of proton-proton ($pp$) collisions collected with the LHCb detector at a centre-of-mass energy of  13\tev between 2016 and 2018, corresponding to an integrated luminosity of 5.4 \invfb.  
In order to avoid experimenter’s bias, the results of the analysis were not examined
until the full procedure had been finalised.

\section{LHCb detector, trigger and simulation}
\label{sec:detector}

The \lhcb detector~\cite{LHCb-DP-2008-001,LHCb-DP-2014-002} is a single-arm forward
spectrometer covering the \mbox{pseudorapidity} $2<\eta<5$,
designed for the study of particles containing \bquark or \cquark
quarks. The detector includes a high-precision charged-particle reconstruction (tracking) 
system consisting of a silicon-strip vertex detector surrounding the $pp$
interaction region~\cite{LHCb-DP-2014-001}, a large-area silicon-strip detector located
upstream of a dipole magnet with a bending power of about
$4{\mathrm{\,T\,m}}$, and three stations of silicon-strip detectors and straw
drift tubes~\cite{LHCb-DP-2013-003,LHCb-DP-2017-001}
placed downstream of the magnet.
The tracking system provides a measurement of the momentum, \ptot, of charged particles with
a relative uncertainty that varies from 0.5\% at low momentum to 1.0\% at 200\gevc.
The minimum distance of a track to a primary vertex (PV), the impact parameter (IP), 
is measured with a resolution of $(15+29/\pt)\mum$,
where \pt is the component of the momentum transverse to the beam, in\,\gevc.
Different types of charged hadrons are distinguished from one another using information
from two ring-imaging Cherenkov (RICH)  detectors~\cite{LHCb-DP-2012-003}. 
Photons, electrons and hadrons are identified by a calorimeter system consisting of
scintillating-pad and preshower detectors, an electromagnetic (ECAL) and a hadronic calorimeter. 
Muons are identified by a system composed of alternating layers of iron and multiwire
proportional chambers~\cite{LHCb-DP-2012-002}.

The online event selection is performed by a trigger~\cite{LHCb-DP-2012-004,LHCb-DP-2019-001}, consisting of a hardware stage,
based on information from the calorimeter and muon systems, followed by a software
stage, which performs full event reconstruction. In the hardware stage, selected events 
are required to have at least one high-\pt electron, muon or hadron. In this analysis, signal candidates belong to three mutually exclusive categories depending on whether the events were selected by:  a high-energy electron from the signal candidate (eTOS);  a reconstructed particle not belonging to the signal candidate (TIS);  a reconstructed hadron from the signal candidate (hTOS).\footnote{TOS: Trigger On Signal, TIS: Trigger Independent of Signal} Candidates in each category should not satisfy the requirements of the previous ones.
The software trigger is organised in two stages. The first requires a two-track
secondary vertex with a significant displacement from any PV, and at least one charged particle
with a minimum requirement on the transverse momentum that is also inconsistent with originating from a PV. The second stage is based on a multivariate algorithm~\cite{BBDT,LHCb-PROC-2015-018} exploiting kinematic, geometric, and lepton-identification criteria to select events containing secondary vertices consistent with the decay of a \bquark hadron.

Simulation is used to optimise the selection, determine the signal model for the fit and obtain the selection efficiencies. In the simulation, $pp$ collisions are generated using \pythia 8~\cite{Sjostrand:2007gs} with a specific \lhcb configuration~\cite{LHCb-PROC-2010-056}.  The $B^0$ decay is assumed to proceed according to a phase-space distribution in the kinematic space. 
The $\tau$ decay is simulated using a dedicated model tuned with BaBar data~\cite{PhysRevD.88.093012} from 
the \tauola decay library ~\cite{JADACH1993361}, while the decays of all other
unstable particles are described by \evtgen~\cite{Lange:2001uf}, in which final-state radiation is generated using \photos~\cite{davidson2015photos}.  Finally, the interaction of the generated particles with the detector, and its response, are implemented using the \geant toolkit~\cite{Allison:2006ve, *Agostinelli:2002hh} as described in Ref.~\cite{LHCb-PROC-2011-006}.

\section{Signal candidate mass reconstruction}
\label{sec:mass}

\begin{figure}[tb]
	\centering
		\includegraphics[width=\linewidth]{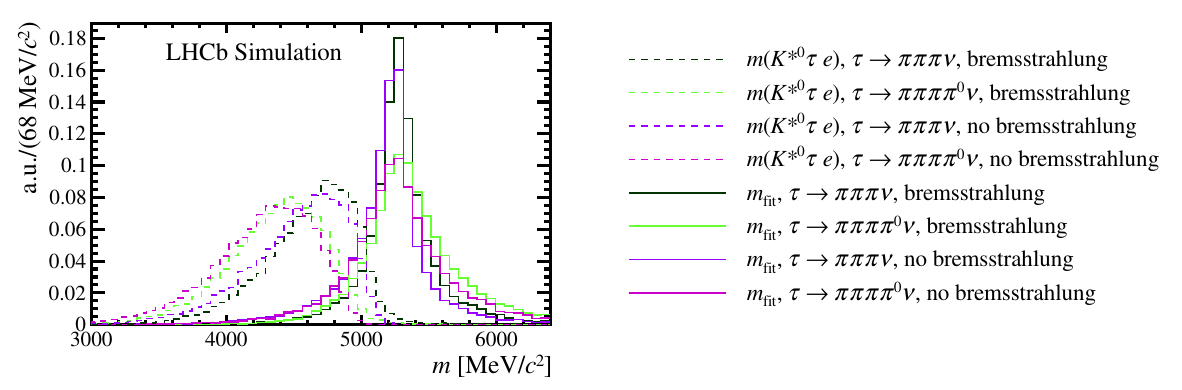}
            \caption{Comparison in simulated signal candidates between (dashed lines) the mass $m(\Kstarz\tau e)$, corresponding to the invariant mass of the tracks in the final state, and (bold lines) the mass $m_{\rm fit}$  accounting for the missing momentum carried from the neutrino via a decay chain fit with kinematic constraints. Legends with the various components of the signal are shown next to the plot. }
	    \label{fig:mass_reco}
\end{figure}

Neutrinos from the $\tau$ decay are not reconstructed since they do not interact with the detector. In addition, the \piz meson from the \tauToPiPiPiPiz decay is ignored during the reconstruction. Hence, the invariant mass of the six tracks in the final state, $m(\Kstarz \tau e)$, peaks at a value lower than the true  $B^0$ meson mass and exhibits degraded resolution~\cite{PDG2024}. Consequently, the  $m(\Kstarz\tau e)$ variable is poorly suited to separate signal from background.   
Instead, the \Bz mass is calculated from the outcome of a kinematic fit in which the entire decay chain is reconstructed imposing the following constraints~\cite{Hulsbergen:2005pu}:  the three charged pions from the tau decay must form a vertex; the missing momentum in the decay chain is attributed to a massless neutrino originating from the $\tau$-decay vertex; the mass of the $\tau$ candidate is constrained to the known $\tau$ mass~\cite{PDG2024}; the $\tau$ candidate must originate from the $B^0$ decay vertex, which is reconstructed using the charged kaon and pion from the \Kstarz decay, along with the electron.  The resulting reconstructed $B^0$ mass, denoted $m_{\rm fit}$, has a largely improved resolution, as shown in Fig.~\ref{fig:mass_reco} for simulated events, and is therefore used in this analysis.

In addition, the electron might emit bremsstrahlung while travelling through the detector. In case this occurs upstream of the magnet, the electron momentum is underestimated and the photons and electron energies are deposited in different calorimeter cells. A dedicated 
procedure searches for energy deposits with transverse energy $\et > 75\mev$, not associated with charged tracks, within a region of the ECAL defined by the extrapolation of the electron track upstream of the magnet. Such “bremsstrahlung clusters” are added to the measured electron momentum, improving its determination. 
The procedure is limited by the energy threshold of the clusters, the acceptance and resolution of the calorimeter, and by the presence of energy deposits wrongly interpreted as bremsstrahlung clusters. 
The bremsstrahlung further worsens the mass resolution, though its impact is less than that due to the missing neutrino in the $\tau$ decay. It is taken into account by modelling the signal shape differently for candidates where bremsstrahlung photons have been added.

\FloatBarrier

\section{Signal candidate selection}
\label{sec:selection}

Candidate $B^0 \to \Kstarz \tau^\pm e^\mp$ decays are reconstructed by combining six charged tracks with $2.0<\eta<4.9$. Five tracks are required to have  $p<110$\gevc and  $\pt>250$\mevc. A sixth track, with  $3<p<110$\gevc and  $\pt>500$\mevc should be within the calorimeter acceptance and satisfy loose electron-identification requirements.  
Candidate \Kstarz mesons are reconstructed by combining two tracks of opposite charge, one compatible with a kaon hypothesis and the other with a pion hypothesis. Those satisfying an invariant mass within $700<m(\Kstarz)<1100$\mevcc undergo a kinematic fit procedure to determine its vertex.
The other three remaining tracks, each identified as pions, with momentum $p>2$\gevc, should originate from another vertex.  
These pions form the $\tau$ candidate, with a charge opposite to that of the electron and with a reconstructed invariant mass  $0.5<m(\tau)<2.0$\gevcc.    
The $\tau$ vertex must have a radial distance between 0.1 and 7.0\mm and a distance along the $z$-axis, pointing along the beam axis, larger than 5.0\mm. Distances are evaluated with respect to the PV associated to the \B candidate.  

The \Kstarz and $\tau$ candidates are required to have a transverse momentum $\pt>1$\gevc. Together with the electron, they form a  $B^0$ candidate, with a decay vertex sufficiently displaced from any  PV.  
Finally, the $\Kstarz\tau$ invariant mass is required to be lower than 5\gevcc and the reconstructed $B^0$ candidate mass $2<m(\Kstarz\tau e)<10$\gevcc. 
A sample of data with $\tau$ and electron of the same charge (referred to as the same-sign sample in the following) is also selected as a proxy for backgrounds.

Background contributions can be divided into two categories:  combinatorial background, arising from random combinations of tracks; and physics background, due to  $b$-hadron decays partially reconstructed and/or reconstructed with misidentified particles. In particular, a large component of the background involves $D$ mesons with a decay time compatible with that of the $\tau$ lepton and decaying into multiple charged tracks. To optimise the rejection of these backgrounds of a different nature,  a multistage selection procedure is used. The procedure is carried out without involving candidates in the region where the signal is expected, referred to as the signal region and defined as $4.6<m_{\rm fit}<6.4$\gevcc. The region $6.4<m_{\rm fit}<12.0$\gevcc is referred to as the upper mass sideband, while the region  $2.0<m_{\rm fit}<4.6$\gevcc as the lower mass sideband. The former is dominated by combinatorial background, while the latter contains both combinatorial and physics background.  
 
The Punzi figure of merit~\cite{Punzi:2003bu},  defined as $\varepsilon/(\sigma/2+{\sqrt{Y_{\rm b}}})$ with $\sigma=3$, is used to optimise the threshold requirement for each selection stage, apart from specific vetoes discussed later. Here, $\varepsilon$ is the efficiency of the signal selection, obtained from simulation, and  $Y_b$ is the background yield in the signal region, estimated differently for each selection stage.

The first stage of the selection is based on a multivariate discriminant exploiting the differences between the topologies of the signal decays and the combinatorial background. A  gradient-boosted decision tree~(GBDT) algorithm~\cite{Friedman} implemented in the TMVA package~\cite{Speckmayer:2010zz} combines information from: the vertex-fit $\chi^2$ of the \Bd and $\tau$ candidates; the \Bd and \Kstarz flight distances; the $\chi^2$ of the flight distances of the \Bd, \Kstarz and the $\tau$ candidates; and the \Bd impact parameter. The \Bd flight distance is evaluated with respect to the PV, while that of the \Kstarz and $\tau$ are evaluated with respect to the \Bd vertex.  
The classifier is trained using simulated samples as a signal proxy and the upper mass sideband of the opposite-sign data sample as a background proxy.  
A \kfolding\ approach~\cite{Bevan:2017stx} is used to exploit the training samples without biasing the output of the classifier.  For the threshold optimisation, the background yield in the signal region is extrapolated from a fit to the $m_{\rm fit}$ distribution of the \Bd candidates in sidebands, modelled by a gamma function. 
The obtained background yield is then corrected to account for possible extrapolation biases by a fit to the same-sign sample in the signal region. 

A particle selected from a partially reconstructed decay is often surrounded by other particles that are not used to reconstruct the candidate. 
To exploit this feature, the second stage of the selection makes use of a Fisher discriminant~\cite{fisher36lda}, built using the \kfolding\ procedure and based on the following variables evaluating the particle isolation: 
the logarithm of the smallest variation of the \Bd  vertex-fit $\chi^2$ when adding  one or two tracks independently; 
the invariant mass of  the \Bd candidate and of the additional track giving the smallest variation to the \Bd  vertex-fit $\chi^2$;
the invariant mass of  the \Kstarz or $\tau$ candidate and of the two additional tracks giving the smallest variation to the \Kstarz or $\tau$  vertex-fit $\chi^2$;
the number of tracks that modify the vertex-fit $\chi^2$ by less than 9 units if added to the \Kstarz vertex; the asymmetry of the transverse momentum of all charged tracks (or, separately, neutral particles) in a cone with aperture $\Delta R= \sqrt{\Delta \phi^2+\Delta \eta^2}=0.4$ around the \Kstarz or $\tau$ direction, where $\phi$ is the polar angle. 
The Fisher discriminant is chosen for its stability in the training procedure, despite the small size of the training samples available at this point in the selection. Signal simulation and, for the background, same-sign data that fall into the signal mass region are used for training and optimisation of the threshold.
The input variables used in both classifiers are validated by checking the agreement between data and simulation on the \BdToDDs sample.

The third stage makes use of the multivariate selection dedicated to the rejection of charmed mesons misidentified as $\tau$ leptons, as originally developed for the \osBToKstTauM search~\cite{LHCb-PAPER-2022-021}.  The $\tau$ lepton predominantly decays into three charged pions and a neutrino through the $a_1^\pm$ resonance, which in turn decays into $\rho^0$ and $\pi^\pm$ mesons. In order to exploit the kinematic properties of this decay chain,  the multivariate selection uses the minimum and maximum of the momenta of the pions from the $\tau$ candidate and the masses of pion pairs with zero combined charge. 
The threshold optimisation follows the same approach used for the first multivariate selection. 

 The fourth selection step consists of particle-identification~(PID) requirements on the final-state particles. Three PID variables, each related to the final-state particle species (electron, kaon and pion) are used.  
 A three-dimensional scan over the possible thresholds of the three particle-identification criteria is performed and a configuration maximising the Punzi figure of merit is chosen. 
For this optimisation, the background yields evaluated in the previous step are multiplied by the rejection of the particle-identification requirements, as estimated from same-sign events within the signal region. 

  Further background is removed by exploiting the discriminating power of the \Kstarz candidates' invariant mass, required to be within $842<m(\Kstarz)<970$\mevcc as a result of an optimisation using the Punzi figure of merit. 

Additionally, selection criteria are imposed on the logarithms of the tau flight distance from the \Bz vertex and its significance 
to eliminate residual physics backgrounds. These requirements have no impact on the events in the selected sidebands, but they reduce additional background components that are present in the control regions used for the background parameterisation.

The remaining physics backgrounds are studied with candidates rejected by the first multivariate selection within the signal region.  
Mass distributions from all possible final-state particle combinations, including particle misidentification, are analysed, and vetoes are imposed to remove the observed resonant structures. For example, \mbox{$B^0 \to D^{*-} e^+ \nu_{e}$} decays, with \mbox{$D^{*-}\to \Dzb \pi^-$} and \mbox{$ \Dzb  \to K^+ \pi^- \pi^+ \pi^-$} decays and analogous contributions from the $D_0^*(2300)$ and $D_2^*(2460)$ resonances, are a background for the \SC mode if three of the four pions are misreconstructed as a $\tau$ candidate and the remaining pion and kaon mimic a  \Kstarz candidate. A similar background comes from  $B^0 \to D_1(2420)^{-}  e^+ \nu_{e}$ decays with $D_1(2420)^{-} \to D^{-} \pi^+ \pi^-$ and $D^{-}\to K^+ \pi^- \pi^-$. 
These two categories of decays are background also for the \OC mode, though, for charge conservation, this additionally requires misidentification of both a pion and an electron.   
To reduce these backgrounds from charmed mesons, candidates with $m(K^+ \tau^-)$ or $m(\Kstarz\pi^-\pi^+)$ within $\pm60\mevcc$ of the known \Dz mass~\cite{PDG2024}  and candidates with $m(\Kstarz\pi^-)$ or $m(K^{+}\pi^-\pi^-)$ within $\pm60\mevcc$ of the known $D^-$ mass~\cite{PDG2024} are removed. Additionally, for the \OC mode, candidates are excluded if the invariant mass of $\Kstarz e^-$ or $K^{+}\pi^-e^-$ combinations, calculated by assigning the pion mass to the electron, lies within $\pm60\mevcc$ of the known $D^-$ mass~\cite{PDG2024}. No other significant contributions from decays with charmed mesons are observed. 
For the \OC  mode,   $B^0 \to \Dstarm \tau^+ \neut$ decays with $\Dstarm\to \Dzb \pim$,  $ \Dzb \to K^+ e^- \neueb$ and $\tau^+ \to \pi^+ \pi^- \pi^+ \neutb$ decays constitute an irreducible background. This background does not peak in the invariant-mass distribution of the kaon and electron due to the missing energy carried by neutrinos. Thus, its contribution is assumed to be absorbed in the background shape parameterisation used in the fit.   

After the selection procedure described above, there is never more than one $B^0$ candidate selected per event.

\section{Normalisation channel}
\label{sec:normalisation}
The normalisation channel  \BdToDDs, with \DToKPiPi and \DspToKKPi decays, is reconstructed using six tracks, each with momentum $2<p<110$\gevc, $\pt>250$\mevc, pseudorapidity in range $2.0<\eta<4.9$ and PID information from the RICH detectors.  Three of these tracks, one identified as a kaon and the other two as pions, originating from a common vertex, displaced from the PV, are used to form a $D^-$ candidate. An invariant mass within $\pm40\mevcc$ of the known $D^-$ mass~\cite{PDG2024} is also required.  
Analogously, one track that is identified as a pion, and two tracks identified as kaons, are used to form a $\Ds$ candidate with a displaced vertex and an invariant mass within $\pm40\mevcc$ of the known $D_s^+$ mass~\cite{PDG2024}.
The same particle-identification requirements used for selecting the signal are applied to the kaon and pion candidates.

\begin{figure}[tb]
	\centering
		\includegraphics[width= .8\linewidth]{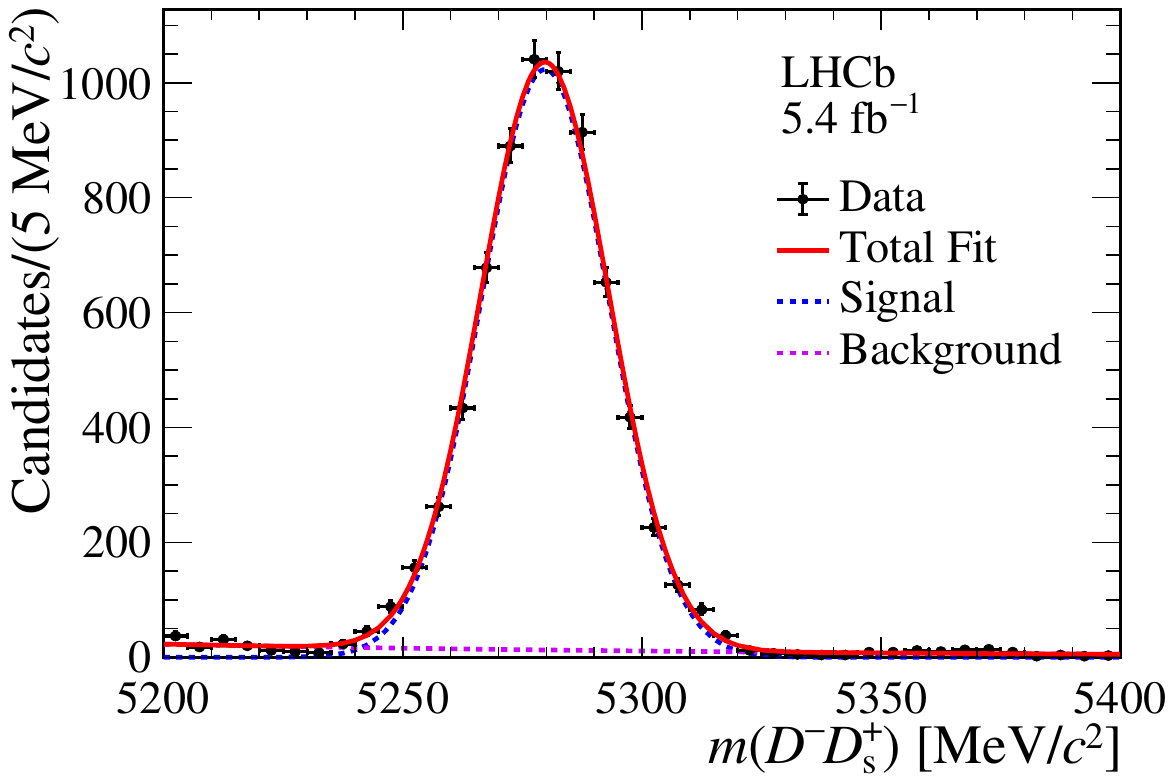}
	    \caption[Normalisation channel fit]{Invariant-mass distribution of selected $D^- D_s^+$ candidates along with the fit result.}
	    \label{fig:dds_fit}
\end{figure}

The normalisation channel yields are measured from unbinned maximum-likelihood fits of the \Bd-candidate invariant-mass distribution in the range \mbox{$5200<m(D^{-}D_s^+)<5400$\mevcc}, separately for each year of data taking. The fit model consists of a Gaussian function to parameterise the signal component and an exponential function to describe the background.  
The fits lead to yields of $2\,054\pm 45$, $2\,086\pm 45$ and $2\,730\pm 52$ events for the 2016, 2017 and 2018 datasets, respectively. 
The mass distribution of the normalisation candidates for the combined dataset, along with the fit results, is shown in Fig.~\ref{fig:dds_fit}.

\section{Determination of efficiencies}
\label{sec:effs}

The efficiencies for the different steps of the selection chain are evaluated separately for each year of data taking and for the \OC and \SC decay channels.  
With the exception of the particle-identification efficiency and the normalisation-channel hardware trigger efficiency, which are directly estimated from data, all efficiencies are estimated from simulation. Electron reconstruction and hardware trigger efficiencies, after being estimated with simulation, are corrected via calibration samples. 
The reliability of the simulation 
is assessed using \Bd, \Dm and \Ds candidates from the normalisation channel \BdToDDs as proxies for two- and three-prong particle vertices, selected with the additional requirement that the \Bd-candidate mass lies within  $\pm 50$\mevcc of its known value~\cite{PDG2024}.
All the variables used in the multivariate selections are reasonably well described by simulation, with the impact of residual discrepancies assessed as a systematic uncertainty.

The corrections to the electron reconstruction efficiencies are evaluated by comparing the reconstruction efficiencies from simulation and data in \mbox{$B^+ \to $\jpsi$ (\to e^+ e^-) K^+$} decays, estimated using the ``tag and probe'' method as documented in Ref.~\cite{LHCb-DP-2019-003}. The corrections are evaluated per year of data taking as a function of the electron \pt, $\eta$ and  $\phi$. They are estimated between 1 and 2\%.  

Hardware trigger efficiencies are calibrated by comparing the reconstruction efficiencies from simulation and data in $B^+ \to$ \jpsi $(\to e^+ e^-) K^+$ decays for the eTOS category and \BdToDDs decays for the TIS and hTOS categories~\cite{LHCb-PUB-2014-039}. Corrections are estimated as a function of relevant variables related to the particle that triggered the event:  electron  \et and electromagnetic-calorimeter granularity for eTOS;  hadron  \et and hadronic-calorimeter granularity for hTOS; \pt of the $B^0$ meson for TIS, which accounts for the correlation with the other $b$ hadron in the event. These corrections lead to variations of up to $\sim15$\% on the hardware trigger efficiencies estimated from simulation.
For the normalisation channel,  the efficiency of the hardware hadron trigger is evaluated directly from data as the fraction of events triggered by both any muon and one of the hadrons from the  \BdToDDs decay, relative to those triggered by any muon alone.

The particle-identification efficiency is evaluated using data calibration samples, separately for each year of data taking.  
A pure sample of pions and kaons is obtained from the decay \mbox{$\decay{\Dstar^+}{\Dz (\to{\kaon\pion})\pion^+}$}, while a pure sample of electrons is obtained from \mbox{$B^+ \to \jpsi (\to e^+ e^-) K^+$} decays, without relying on PID selection criteria~\cite{LHCb-DP-2018-001}.  The efficiency of the PID requirements is then computed in intervals of the kinematic variables of the final-state particle momentum and pseudorapidity~\cite{LHCb-PUB-2016-021}. Intervals are chosen to balance the stability of the efficiency within each interval and the statistical uncertainty.  The event occupancy, parameterised by the number of tracks per event, is also considered,  relying on observed distributions from same-sign events in the case of signal, and the data distribution itself for the normalisation channel. 

The reconstruction and selection efficiency is determined from simulation, assuming a phase-space model for signal, \ie, a uniform distribution of candidates as a function of the angular observables.    
The efficiency is not uniform as function of the relevant kinematic observables, which are the four-momentum-transfer $q^2$ and the three angular observables $\varphi$, $\cos(\theta_l)$ and $\cos(\theta_K)$, defined as in Ref.~\cite{LHCb-PAPER-2013-019}.   The distributions of the efficiency as a function of these variables are provided in Appendix~\ref{App:A}.  These distributions can be used to interpret the results  for models having \SC and \OC decay kinematics different from the phase-space model used in this search.   

\section{Fit strategy}
\label{sec:fitMass}
An extended unbinned maximum-likelihood fit is performed independently on the  $m_{\rm fit}$ distributions for the \OC and \SC decay channels.  For each channel the distribution is described by a 
function $P_{\rm{tot}}$, which is the sum  of three components,   $P_{\rm{tot}} =  Y_{3\pi}  P_{3\pi} +  Y_{3\pi\pi^0}  P_{3\pi\pi^0} + Y_{\rm{bkg}}  P_{\rm{bkg}}$.  The function $P_{3\pi}$ models the dominant signal contribution with \tauToPiPiPi 
and $Y_{3\pi}$ indicates the corresponding yield. Analogously,  $P_{3\pi\pi^0}$  models the subdominant  signal contribution with  \tauToPiPiPiPiz 
and  $Y_{3\pi\pi^0}$ represents the corresponding yield.  Finally, $P_{\rm{bkg}}$ models the background, with yield parameter $Y_{\rm{bkg}}$.   

The signal components $P_{3\pi}$ and   $P_{3\pi\pi^0}$ are parameterised independently, as they are expected to have slightly different shapes. For each of the components, the sum of two double-sided Crystal Ball functions (DSCB) are used,  accounting for candidates with and without bremsstrahlung  recovery. The fraction of candidates with bremsstrahlung recovery is fixed from simulation.  A DSCB is a Crystal Ball function~\cite{Skwarnicki:1986xj} with an additional power-law tail on the right side of the peak.  The parameters are determined and fixed from fits to the simulated signal samples.    
The  yields for the dominant signal contribution are expressed for the \OC channel as 
\begin{equation}
Y_{3\pi} =   
\frac{ \BR(\OC)   \BR(\KstToKPi)   \BR (\tauToPiPiPi)}{  \BR(\BdToDDs)  \BR(\DToKPiPi)   \BR(\DspToKKPi) }  
\sum_{i \in {\rm{year}}}
{\left(\frac{\varepsilon   Y_{\rm{N}}  }{\varepsilon_{\rm{N}}}\right)_{i}}.
\end{equation}
 Analogous expressions hold for the \SC channel and for the subdominant signal components. Here, $\epsilon$ represents the signal efficiency, while $Y_{\rm{N}}$ and $\epsilon_{\rm{N}}$ indicate the yield and efficiency for the normalisation channel. The ratio $\varepsilon Y_{\rm{N}}  /\varepsilon_{\rm{N}}$ is calculated for each year, as indicated by the $i$ index. The branching fractions are taken from Ref.~\cite{PDG2024}, with the exception of  the parameter of interest $\BR(\OC)$, common among the yields of the dominant and subdominant signal contributions,  and determined from the simultaneous fit. A bias on the measured signal branching ratio, at the level of $10^{-7}$, is evaluated with background-only pseudoexperiments and subtracted.  All other parameters are constrained with Gaussian functions to account for the systematic uncertainties described in the next section. 

An empirical functional form is used to parametrise the background shape. Based on a control sample obtained by inverting the combinatorial multivariate selection on the selected \OC and \SC candidates, a DSCB function is chosen.  The shape parameters are constrained to the values determined from the control sample fit, while the background yields are allowed to vary freely.  Signal contamination in the control sample  
is estimated to be less than 5\% of the total signal, assuming a branching fraction of $10^{-5}$.

\section{Systematic uncertainties}
\label{sec:systematics}  
In the ratios of signal and normalisation-channel efficiencies,  systematic uncertainties largely cancel. The uncertainty related to the limited size of the simulation samples used to determine the efficiencies is included. Residual systematic effects in the ratio are assessed as described in the following. 
 
The conversion of the signal yields into their respective branching fractions involves external input from known intermediate branching fractions.  The corresponding systematic uncertainty is assigned by imposing constraints using Gaussian priors from
 Ref.~\cite{PDG2024}.

The systematic uncertainty associated
with the fit model used to measure the normalisation channel yields $Y_n$ is assessed using pseudoexperiments
generated from a kernel-density estimation of the \BdToDDs mass distribution and an exponential function for the background. The pseudoexperiments are subsequently fitted using the baseline model for the normalisation channel. The relative systematic uncertainty is 0.2\%, evaluated as the ratio between the mean fitted and generated yields.

The ratio of the tracking efficiency between the signal and normalisation channels is determined from simulation. Possible differences between data and simulation largely cancel in the ratio. Since the sixth hadron in the normalisation channel interacts differently with the material with respect to the electron in the signal channel, in the tracking efficiency ratios, a 1.4\% uncertainty
is assigned to the hadron~\cite{LHCb-DP-2013-002} and a 0.6\% uncertainty to the electron~\cite{LHCb-DP-2019-003}, evaluated with data control samples.   

The systematic uncertainty in determining the efficiency of the PID selection considers the limited size of the simulation and calibration samples; choice of the bin size of $p$, $\eta$ and the number of tracks; and the use of the \sPlot technique~\cite{Pivk:2004ty} to extract the signal yield in the control samples.

Although the simulation of the variables used in the multivariate classifiers is validated with the normalisation channel, small residual discrepancies could affect the evaluation of the classifier efficiencies. To account for this, the background is statistically subtracted from the \BdToDDs sample candidates by fitting their mass distribution after the selection, as previously described (see Fig.~\ref{fig:dds_fit}). 
The variables of the classifiers relative to the $\tau$ lepton are applied to the \Dm meson and those of the \Kstarz meson are applied to the \Ds meson. For each classifier, the requirement giving an efficiency on the \BdToDDs simulation sample equivalent to that obtained by the requirement on the simulated \SCOC sample is applied. The relative systematic uncertainty is defined as the absolute difference in the efficiencies of this requirement between the \BdToDDs simulation and data, divided by the efficiency in simulation.

The calibration procedure of the hardware trigger efficiency, performed separately for each trigger category, has associated  systematic effects. The first uncertainty is related to the limited size of the data and simulation samples used. In addition, the impact of different binning schemes is checked for the relevant variables used by the calibration. This turns out to be the dominant systematic effect for the hardware trigger efficiencies. For the TIS category, the \pt distribution of the $B$ meson, taken by default from simulation,  is also corrected using the \BdToDDs data sample to assess the impact on the determination of the TIS efficiency. 
The largest difference in the efficiency obtained using data samples alternative to that used in the baseline is also accounted for as an uncertainty. 
Conversely, for the software trigger, no significant discrepancy is observed between the efficiencies determined from simulation and from the high purity \BdToDDs sample. 

Another source of systematic uncertainty is connected to the description of the background model. The parameters of the DSCB function describing the background are constrained using Gaussian priors derived from a fit to a background control region, obtained by loosening the combinatorial multivariate selection. Alternative control regions are defined by choosing different requirements on the combinatorial multivariate discriminant, Fisher discriminant and the \Kstarz mass window, one by one. The peak position and width of the DSCB function are Gaussian-constrained with a width of $\sim 2$\% and $\sim 14$\% respectively, accounting for their variation in these alternative control regions. 

In addition, a systematic uncertainty is assigned to the choice of the DSCB functional form. This is assessed by generating samples via a kernel-density estimation of background control samples and fitting them with the DSCB function. This systematic is added to each of the signal branching fractions, constraining its value with a Gaussian function with a mean of zero and a width set to the average value of the bias measured on the pseudoexperiments, which is at the level of $10^{-7}$.    
The systematic uncertainty on the signal model includes a contribution associated with the reliability of the mass reconstruction technique. It is assessed by reconstructing \BdToDDs candidates with only five tracks, thus emulating the missing energy carried by the neutrino from the $\tau$ decay in the signal channel. The $m_{\rm fit}$ distribution is then fit in the region $5150<m_{\rm fit}<5420$ \mevcc, both for selected simulated and data candidates, parameterising the signal with a DSCB function with symmetric widths and the background with an exponential function. Discrepancies between data and simulation on the peak position and the widths of the DSCB are summed in quadrature, with the statistical uncertainties coming from the determination of their values from simulation. The results are used to constrain using Gaussian prior the fit to the data. Additionally, the signal-model systematic uncertainty has a component related to the fraction of bremsstrahlung candidates taken from the simulation. The statistical uncertainty of the simulation sample is combined with the largest discrepancy between the fraction of bremsstrahlung candidates in data and simulation evaluated in $B^0\to \Kstarz\jpsi(\to e^+ e^-)$ decays, as described  in Ref.~\cite{LHCb-PAPER-2022-045}. 

The increase of the observed upper limit when progressively adding each of the systematic uncertainties is shown in Table~\ref{tab:syst}. 
The dominant systematic effect comes from the uncertainty on the arbitrary choice of the background control region. 

\begin{table}[t]
	\centering
	\caption{Relative increase of the observed upper limit when applying each systematic uncertainty in turn. Each number is expressed as a function of the previous step, while the total is relative to the case when no systematic is included.}
	\label{tab:syst}
	\renewcommand\arraystretch{1.3}	\begin{tabular}{l|cc}
		Systematic effect & \multicolumn{2}{c}{Upper limit increase [\%]} \\
		 &  \textbf{\OC} & \textbf{\SC} \\
 
		\hline
         Input branching fractions & ~~2.3 & ~~2.5 \\
         Normalisation yields & $<$0.1 & $<$0.1 \\
         Efficiencies & ~~1.2 & ~~1.0 \\
         Background model & ~~4.7 & ~~5.2 \\
         Signal model & ~~1.2 & ~~0.5 \\
         \hline
         Total & ~~9.7 & ~~9.5 \\
	\end{tabular}
\end{table}

\section{Results}
\label{sec:results}

\begin{figure}[htbp]
	\centering
    \includegraphics[width= .43\linewidth]{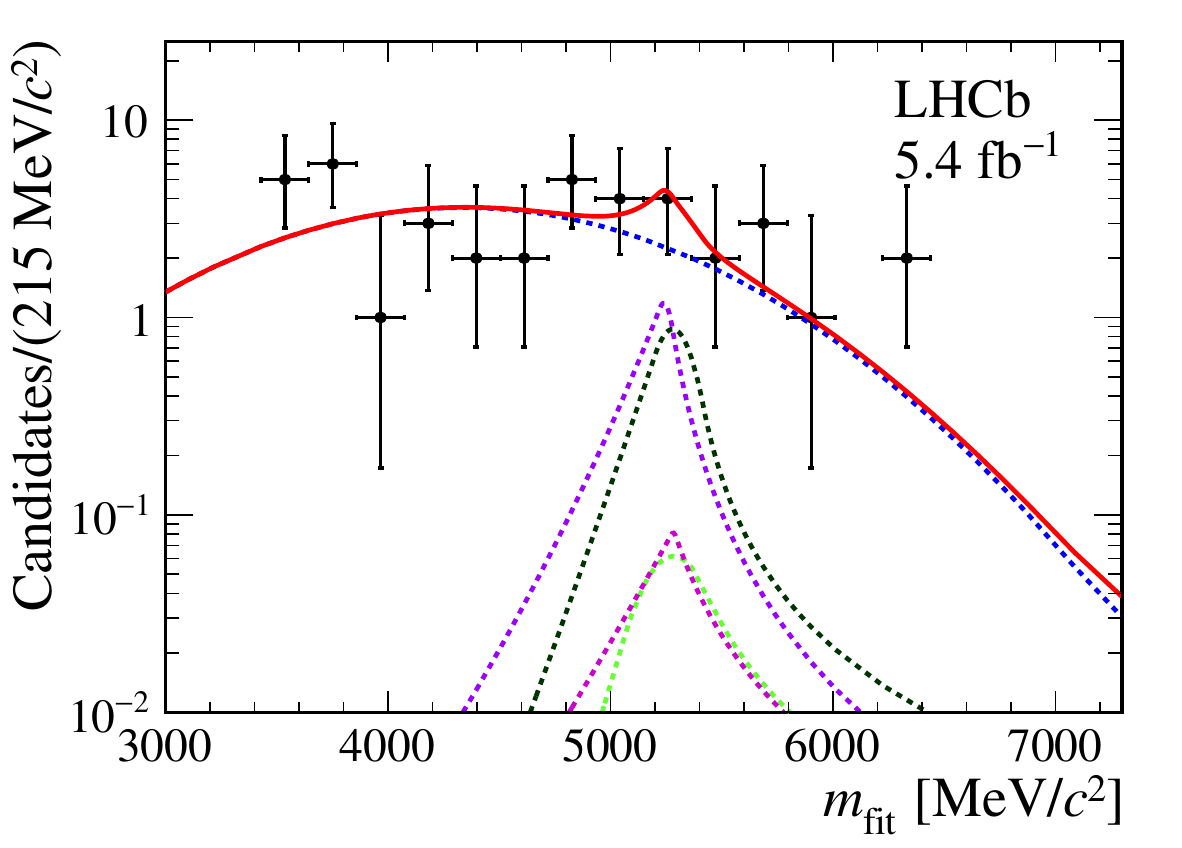}
	\includegraphics[width= .55\linewidth]{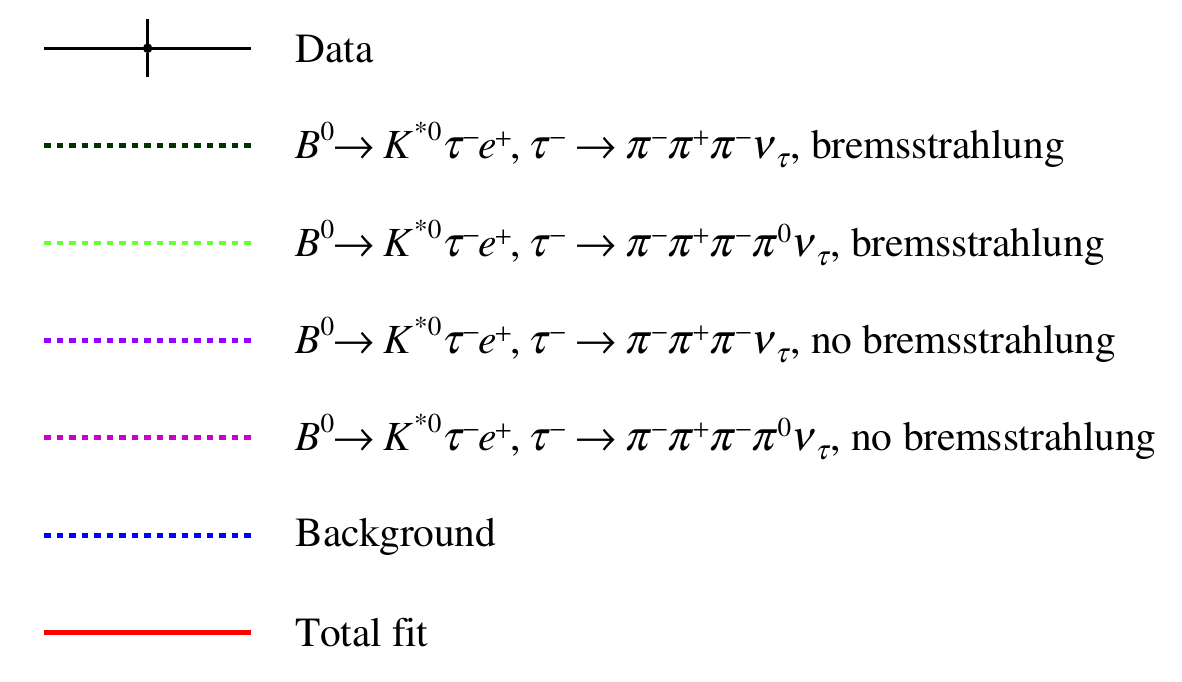}
	\includegraphics[width= .43\linewidth]{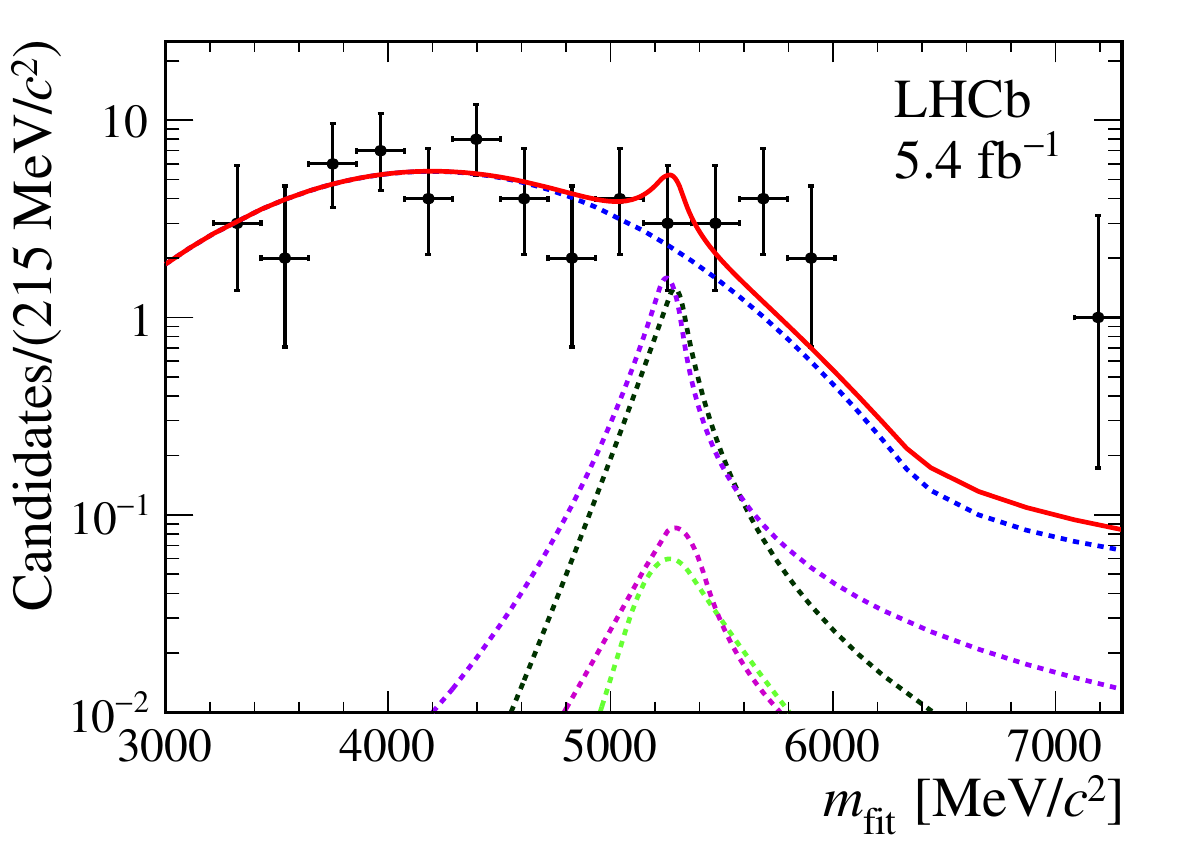}
	\includegraphics[width= .55\linewidth]{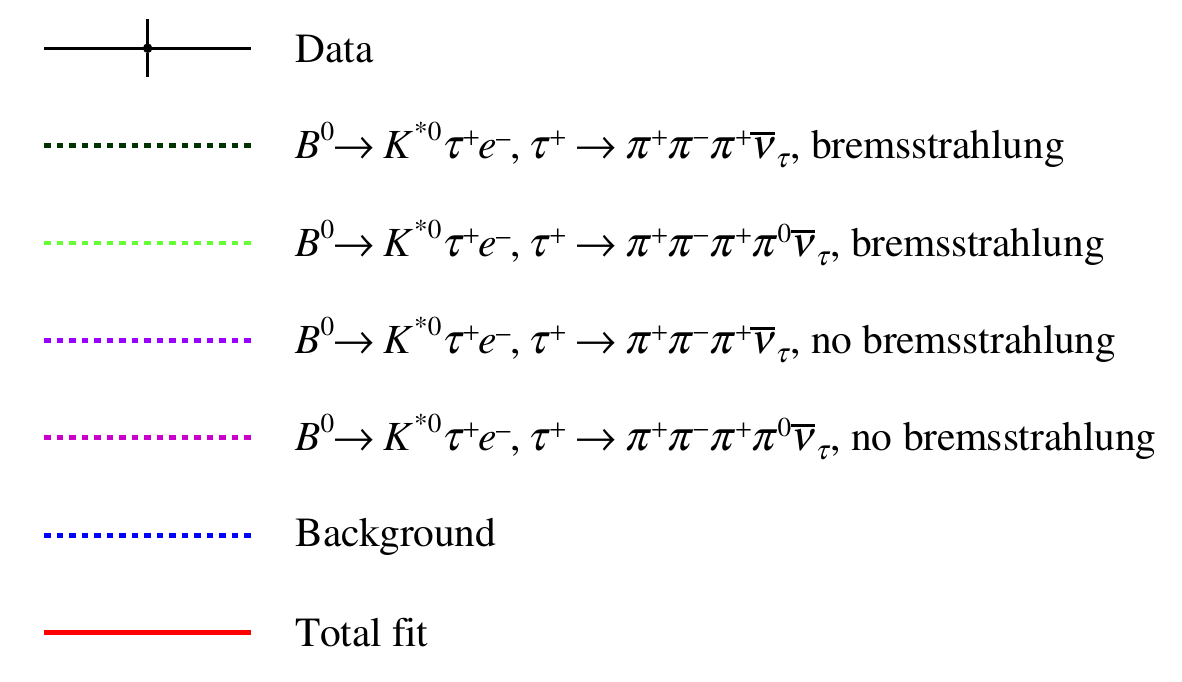}
	    \caption[Fit]{
	    Distribution of the mass $m_{\rm fit}$  of  (top) \SC  and (bottom) \OC  selected candidates, with the simultaneous fit result. Legends with the various components of the fit are shown next to the plots.
	    }
	    \label{fig:fitResults}
\end{figure}

\begin{figure}[htbp]
	\centering
	\includegraphics[width= .48\linewidth]{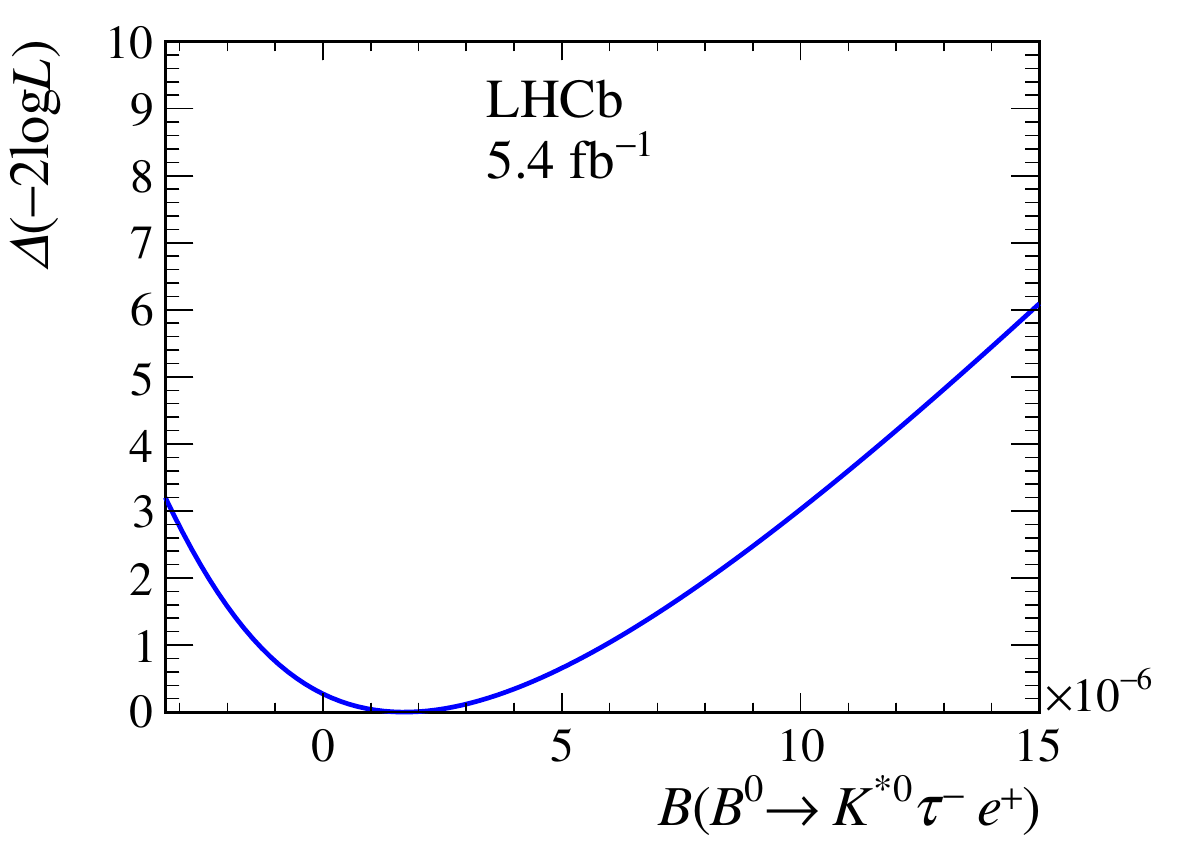}
	\includegraphics[width= .48\linewidth]{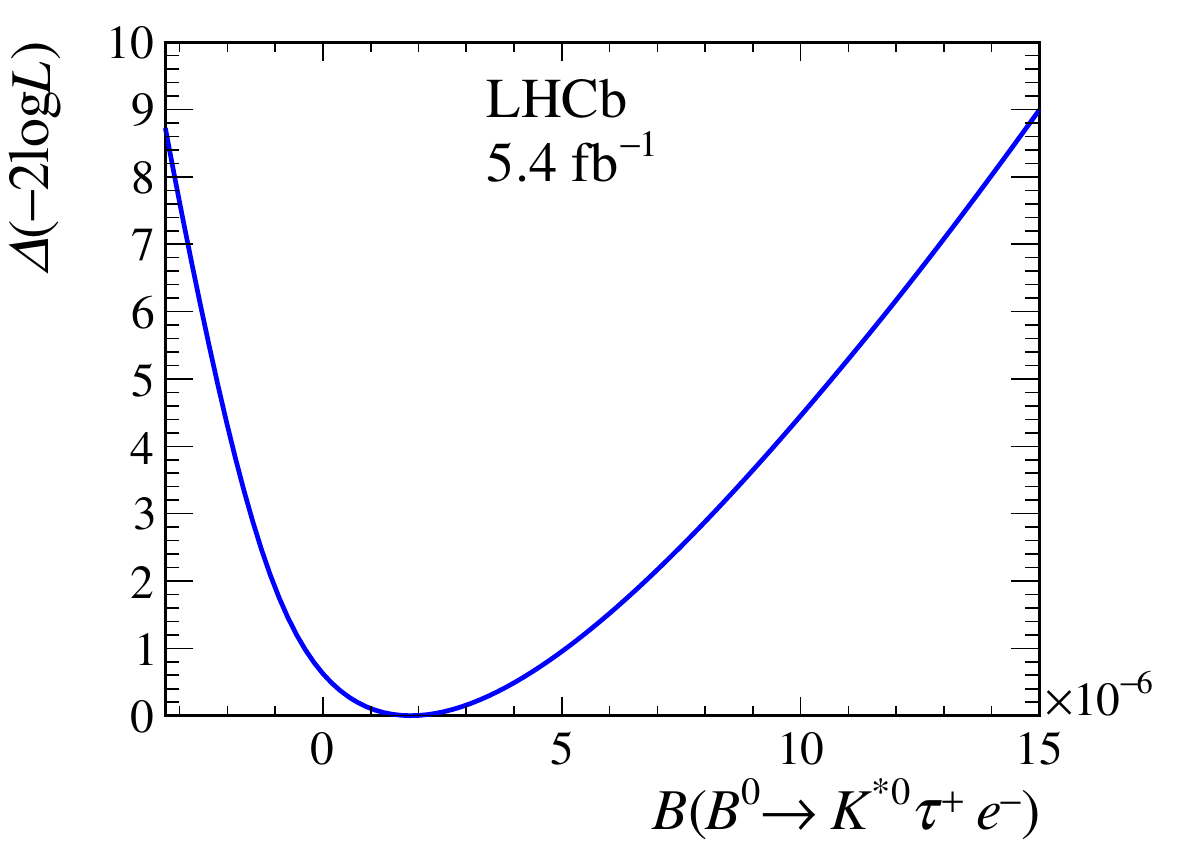}
	    \caption[Profile likelihoods]{Likelihood-ratio scans for the  (left) \SC  and (right) \OC  branching fraction with all nuisance parameters profiled. }
	    \label{fig:profilelikelihoods}
\end{figure}

\begin{figure}[htbp]
	\centering
	\includegraphics[width= .48\linewidth]{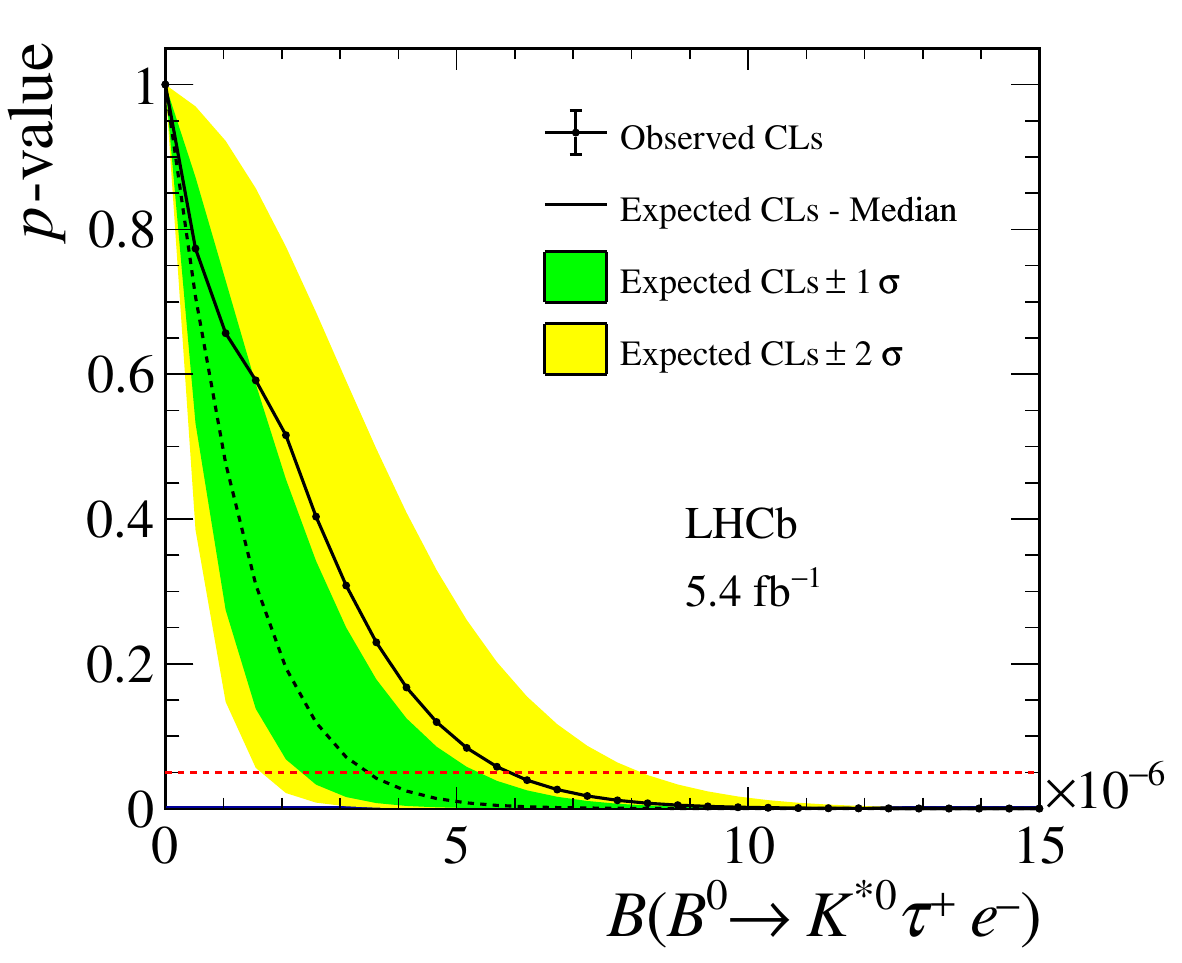}
	\includegraphics[width= .48\linewidth]{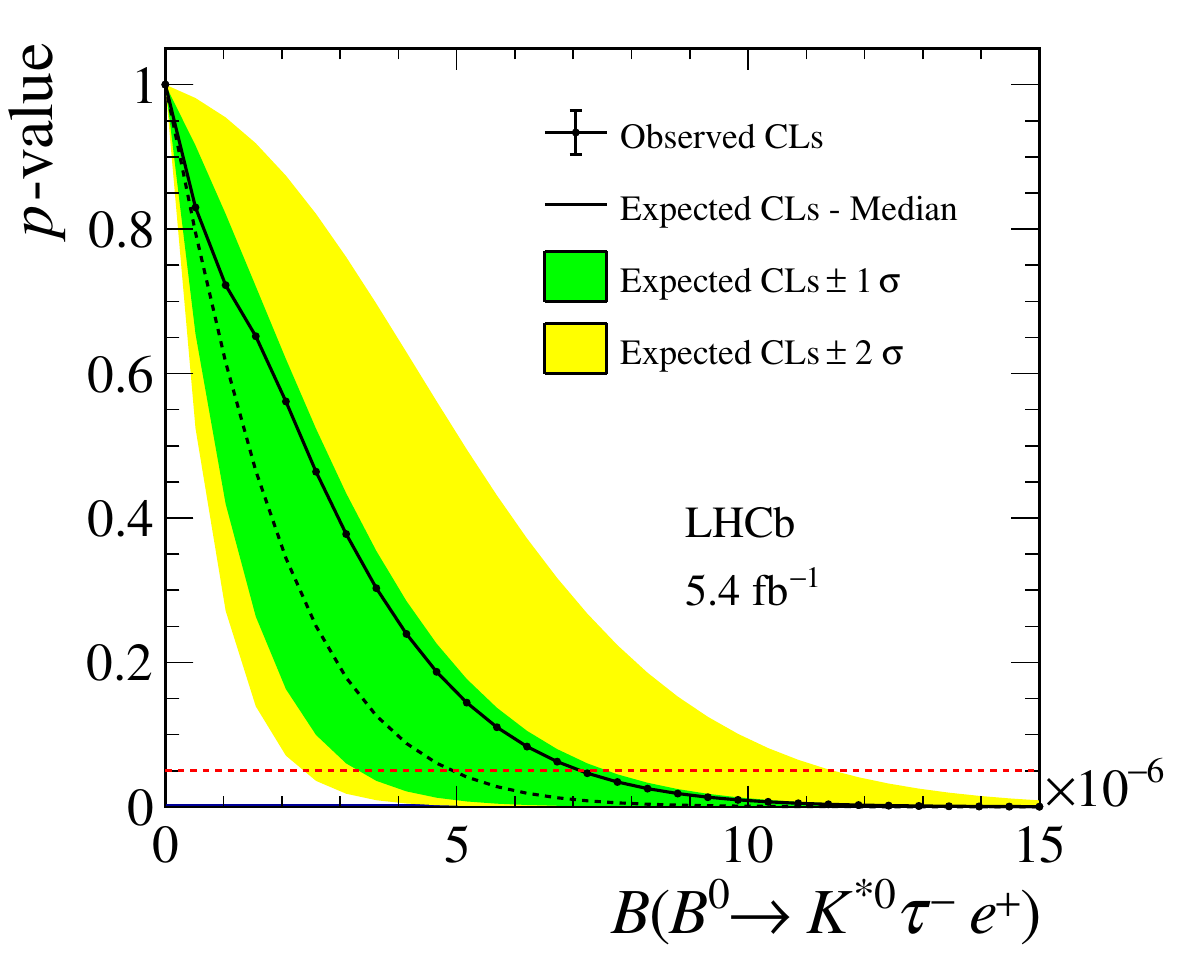}
	    \caption[Limits]{The expected and observed $p$-values derived with the $\mathrm{CL_s}$ method as a function of  the  (left) \SC  and (right) \OC  branching fraction. The red line corresponds to the 95\% CL.  }
	    \label{fig:limits}
\end{figure}

The results of the extended unbinned maximum-likelihood fit to the $m_{\rm fit}$ distributions are shown in Fig.~\ref{fig:fitResults}, with the corresponding likelihood-ratio scans, with all nuisance parameters profiled, shown in Fig.~\ref{fig:profilelikelihoods}.  The likelihood ratio is defined as the likelihood of the hypothesis with no signal over the likelihood of the hypothesis of a nonzero branching fraction. No significant signal contribution is observed and the measured branching fractions correspond to \mbox{$\BR(\SC)= (1.7 \pm 2.5) \times 10^{-6} $} and \mbox{$\BR(\OC)= (1.8 \pm 1.9) \times 10^{-6} $}, accounting for both statistical and systematic sources of uncertainties. Therefore, upper limits on the branching fractions are set  using the $\mathrm{CL_s}$ method with the asymptotic approximation~\cite{CLs, Cowan:2010js}: 
\begin{align} \nonumber \mbox{$\BR(\SC)< 5.9$ $(7.1)\times 10^{-6} $} \end{align} and \begin{align} \nonumber \mbox{$\BR(\OC)< 4.9$ $(5.9) \times 10^{-6}$} \end{align} at the 90\% (95)\% confidence level, as shown in Fig.~\ref{fig:limits}.

These results are obtained assuming a phase-space model for the signal decays.  When computing limits in alternative scenarios, the variation of the reconstruction and selection efficiency with decay kinematics given in Appendix~\ref{App:A}  needs to be accounted for.

As an example, two scenarios of physics beyond the SM are considered in this study: a left-handed model with $C_9^{\tau e} = -C_{10}^{\tau e} 
 \neq 0$ and a scalar model with $C_S^{\tau e} \neq 0$, where $C_i^{\tau e}$ are the lepton-flavour-violating Wilson coefficients~\cite{Becirevic:2024vwy}. For these scenarios, the comparison with phase-space kinematics is shown in Appendix~\ref{app:BSM}. The resulting limits are presented in Table~\ref{tab:NPlimits}.

\begin{table}[t]
	\centering
	\caption{Upper limits on the branching fraction of the decays for the assumed phase-space model and two beyond the SM scenarios, at 90\% (95\%) confidence level.} 
	\label{tab:NPlimits}

	\begin{tabular}{l|cc}
		Model & \multicolumn{2}{c}{Upper limit [$10^{-6}$]} \\ 
        & $\SC$ & $\OC$ \\
 		\hline
       {Phase space (PHSP)} & 5.9 (7.1) & 4.9 (5.9)\\
       {Left-handed ($C_9^{\tau e} = -C_{10}^{\tau e} 
 \neq 0$)} & 6.3 (7.7) & 5.4 (6.4) \\
        {Scalar ($C_S^{\tau e} \neq 0$)} & 6.6 (8.0) & 5.7 (6.8) \\
	\end{tabular}
\end{table}

\FloatBarrier

\section{Conclusion}
\label{sec:conclusion}

This study presents the first search at LHCb for the lepton-flavour-violating decays $B^0 \to \Kstarz \tau^\pm e^\mp$, performed on a sample of proton-proton collision data collected with the LHCb detector at a centre-of-mass energy of  13\tev  between 2016 and 2018, and corresponding to an integrated luminosity of 5.4 \invfb. No significant signal is observed, and upper limits on the branching fractions are set:  \mbox{$\BR(\SC)< 5.9$ $(7.1)\times 10^{-6} $} and \mbox{$\BR(\OC)< 4.9$ $(5.9) \times 10^{-6}$}  at the 90\% (95\%) confidence level. These results assume a uniform distribution of the signal events within the phase space accessible to the \Kstarz, $\tau$, and electron, and are the most stringent upper limits to date on  $b \to s \tau e$ transitions.

\section*{Acknowledgements}
%
%
\noindent We express our gratitude to our colleagues in the CERN
accelerator departments for the excellent performance of the LHC. We
thank the technical and administrative staff at the LHCb
institutes.
We acknowledge support from CERN and from the national agencies:
ARC (Australia);
CAPES, CNPq, FAPERJ and FINEP (Brazil); 
MOST and NSFC (China); 
CNRS/IN2P3 (France); 
BMFTR, DFG and MPG (Germany); 
INFN (Italy); 
NWO (Netherlands); 
MNiSW and NCN (Poland); 
MCID/IFA (Romania); 
MICIU and AEI (Spain);
SNSF and SER (Switzerland); 
NASU (Ukraine); 
STFC (United Kingdom); 
DOE NP and NSF (USA).
We acknowledge the computing resources that are provided by ARDC (Australia), 
CBPF (Brazil),
CERN, 
IHEP and LZU (China),
IN2P3 (France), 
KIT and DESY (Germany), 
INFN (Italy), 
SURF (Netherlands),
Polish WLCG (Poland),
IFIN-HH (Romania), 
PIC (Spain), CSCS (Switzerland), 
and GridPP (United Kingdom).
We are indebted to the communities behind the multiple open-source
software packages on which we depend.
Individual groups or members have received support from
Key Research Program of Frontier Sciences of CAS, CAS PIFI, CAS CCEPP, 
Fundamental Research Funds for the Central Universities,  and Sci.\ \& Tech.\ Program of Guangzhou (China);
Minciencias (Colombia);
EPLANET, Marie Sk\l{}odowska-Curie Actions, ERC and NextGenerationEU (European Union);
A*MIDEX, ANR, IPhU and Labex P2IO, and R\'{e}gion Auvergne-Rh\^{o}ne-Alpes (France);
Alexander-von-Humboldt Foundation (Germany);
ICSC (Italy); 
Severo Ochoa and Mar\'ia de Maeztu Units of Excellence, GVA, XuntaGal, GENCAT, InTalent-Inditex and Prog.~Atracci\'on Talento CM (Spain);
SRC (Sweden);
the Leverhulme Trust, the Royal Society and UKRI (United Kingdom).



\section*{Appendices}

\appendix

\section{Efficiencies as function of kinematics} \label{App:A}

\begin{figure}[htbp]
\centering
\includegraphics[width=\linewidth]{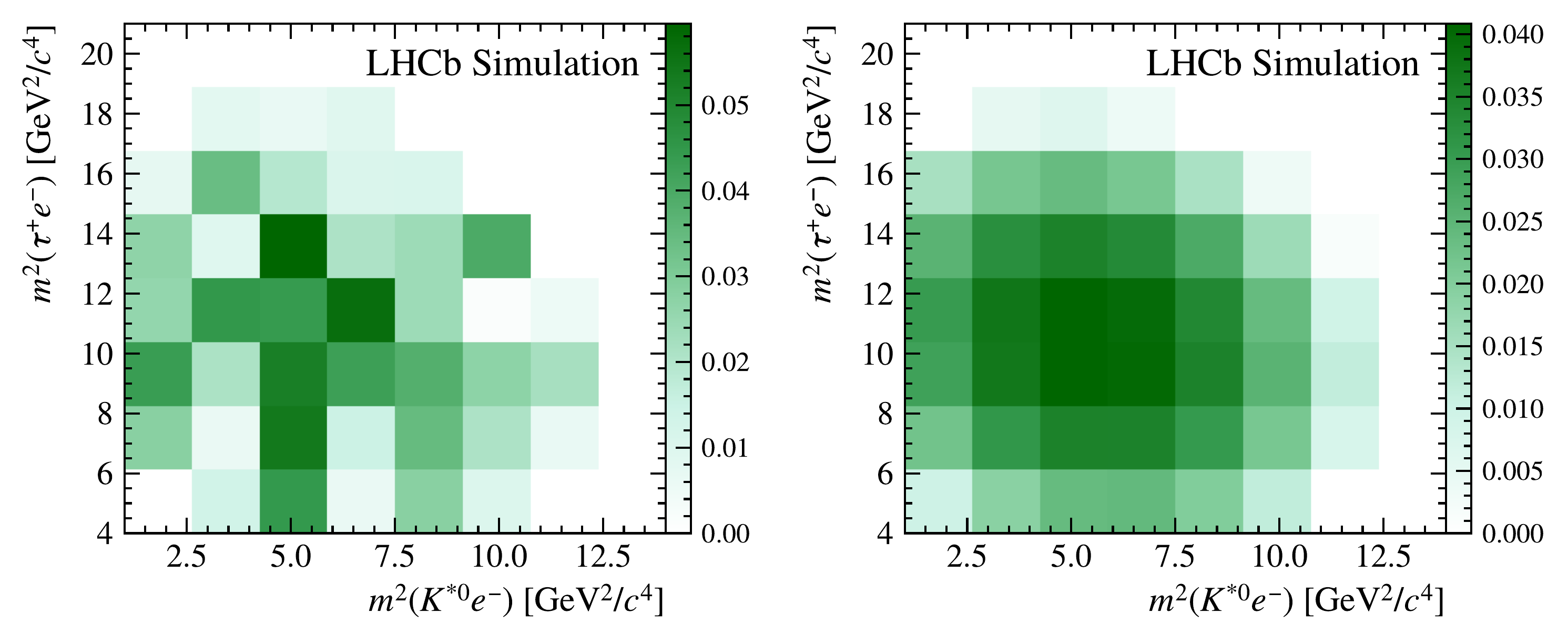}
\includegraphics[width=\linewidth]{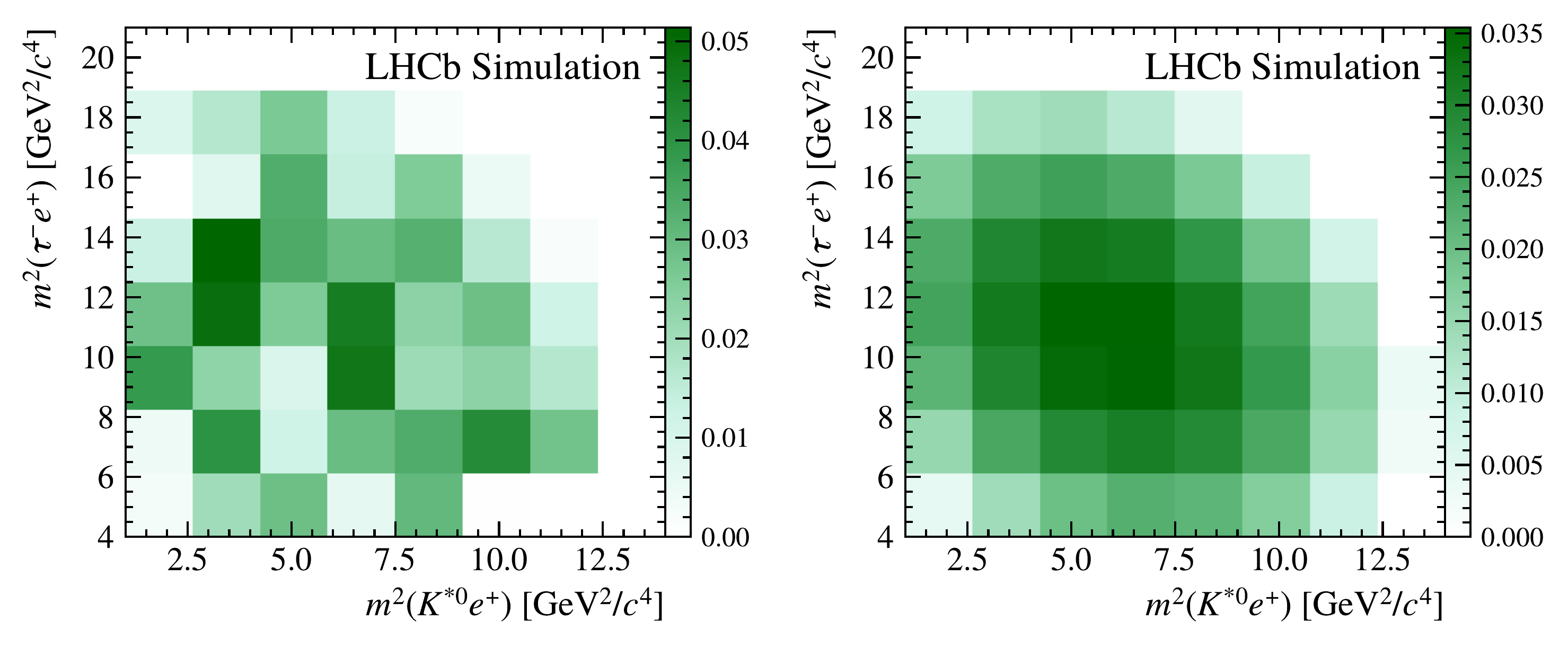}
	    \caption[Distribution of the efficiencies]{Efficiency distribution as a function of~(left) $m^2_{K^{*0} e^{\pm}}$ and $m^2_{\tau^{\mp} e^{\pm}}$ with~(right) their bidimensional polynomial fits for~(top) the \mbox{\OC} and~(bottom) \mbox{\SC} cases, normalised to unity.
     }
	    \label{fig:eff}
\end{figure}

\begin{figure}[htb!]
    \centering
    \includegraphics[width=\linewidth]{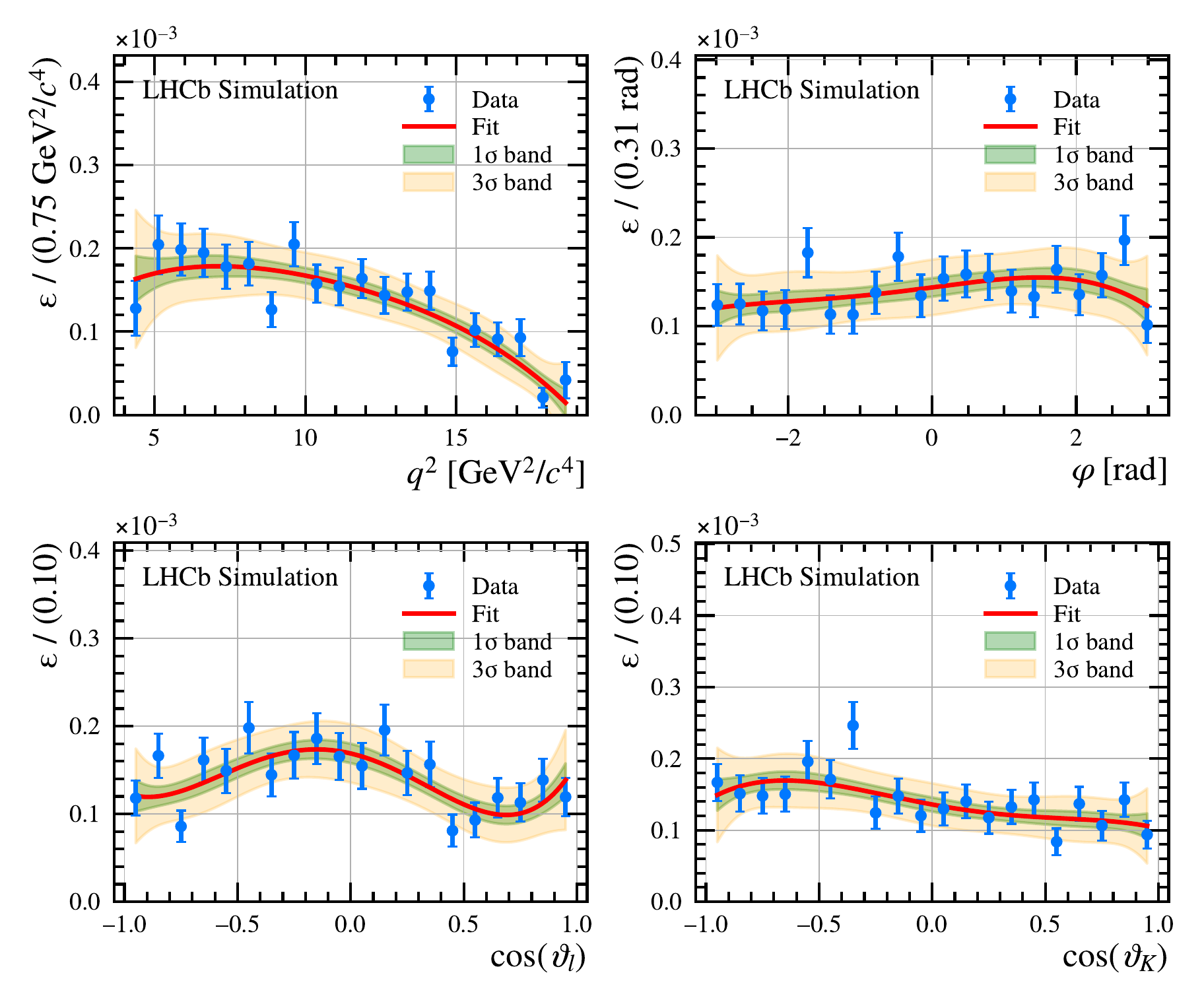}
    \caption{Efficiency distributions with their polynomial fit results as a function of (top left)~$q^2$, (top right)~$\varphi$, (bottom left)~$\cos(\theta_l)$ and (bottom right)~$\cos(\theta_K)$  for \OC and \mbox{$\tau^+ \to \pi^+ \pi^- \pi^+ \bar{\nu_{\tau}}$} decays.}
    \label{fig::np_efficiencies_oc_3pi}
\end{figure}

\begin{figure}[htb!]
    \centering
    \includegraphics[width=\linewidth]{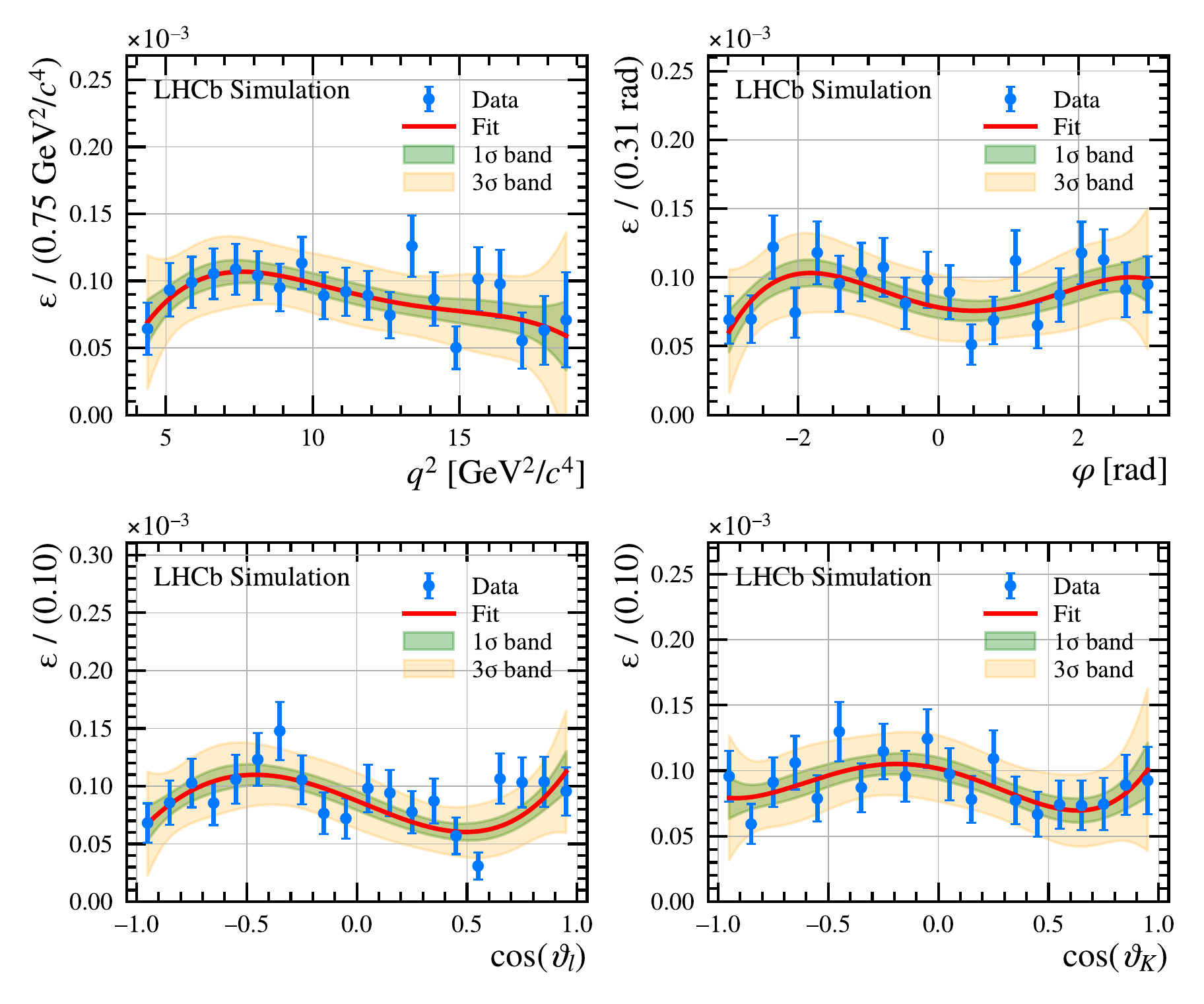}
    \caption{Efficiency distributions with their polynomial fit results as a function of (top left)~$q^2$, (top right)~$\varphi$, (bottom left)~$\cos(\theta_l)$ and (bottom right)~$\cos(\theta_K)$ for \OC and \mbox{$\tau^+ \to \pi^+ \pi^- \pi^+ \pi^0 \bar{\nu_{\tau}}$}   decays.}
    \label{fig::np_efficiencies_oc_3pipi0}
\end{figure}

\begin{figure}[htb!]
    \centering
    \includegraphics[width=\linewidth]{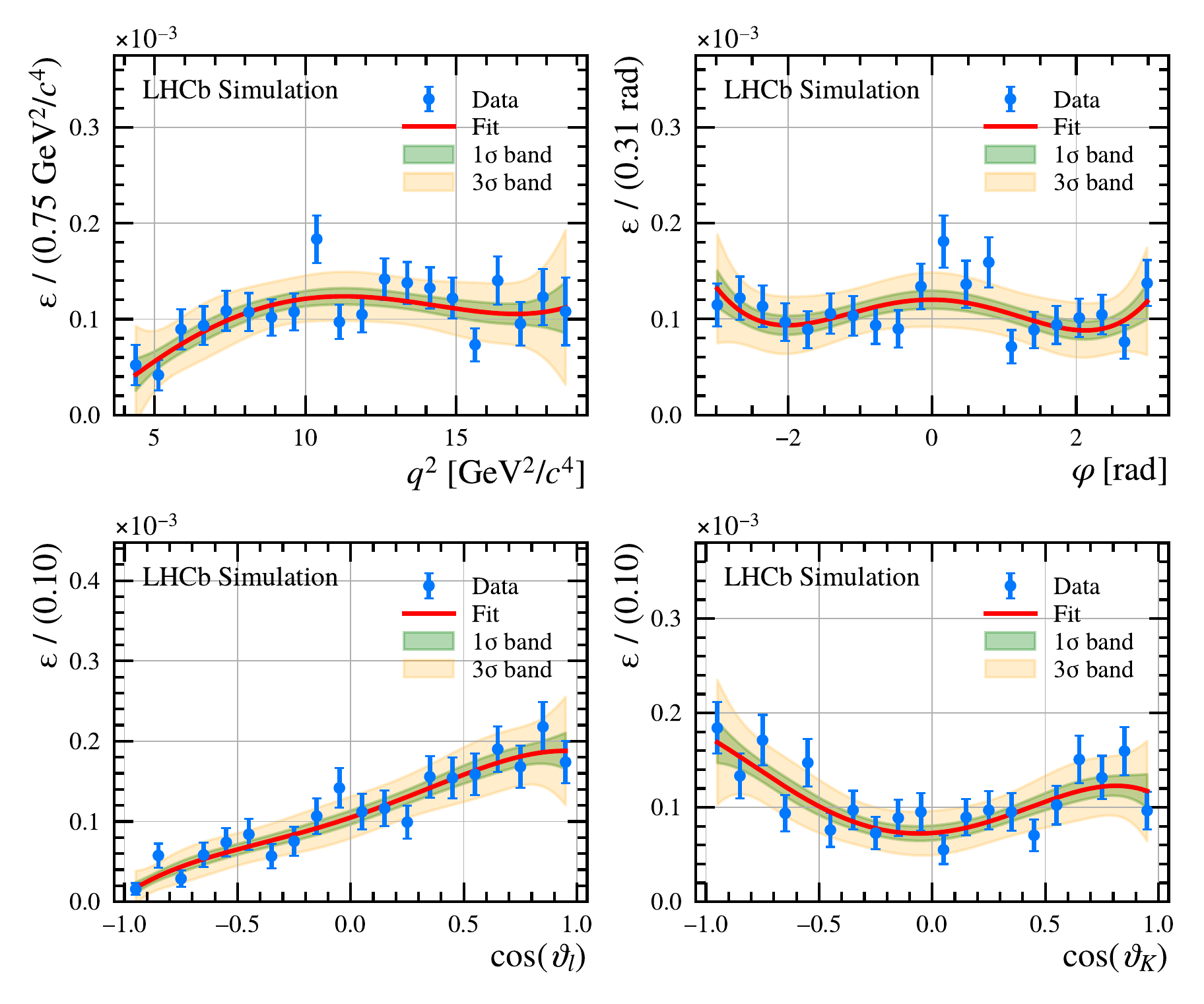}
    \caption{Efficiency distributions with their polynomial fit results as a function of (top left)~$q^2$, (top right)~$\varphi$, (bottom left)~$\cos(\theta_l)$ and (bottom right)~$\cos(\theta_K)$ for \SC and \tauToPiPiPi decays.}
    \label{fig::np_efficiencies_sc_3pi}
\end{figure}

\begin{figure}[htb!]
    \centering
    \includegraphics[width=\linewidth]{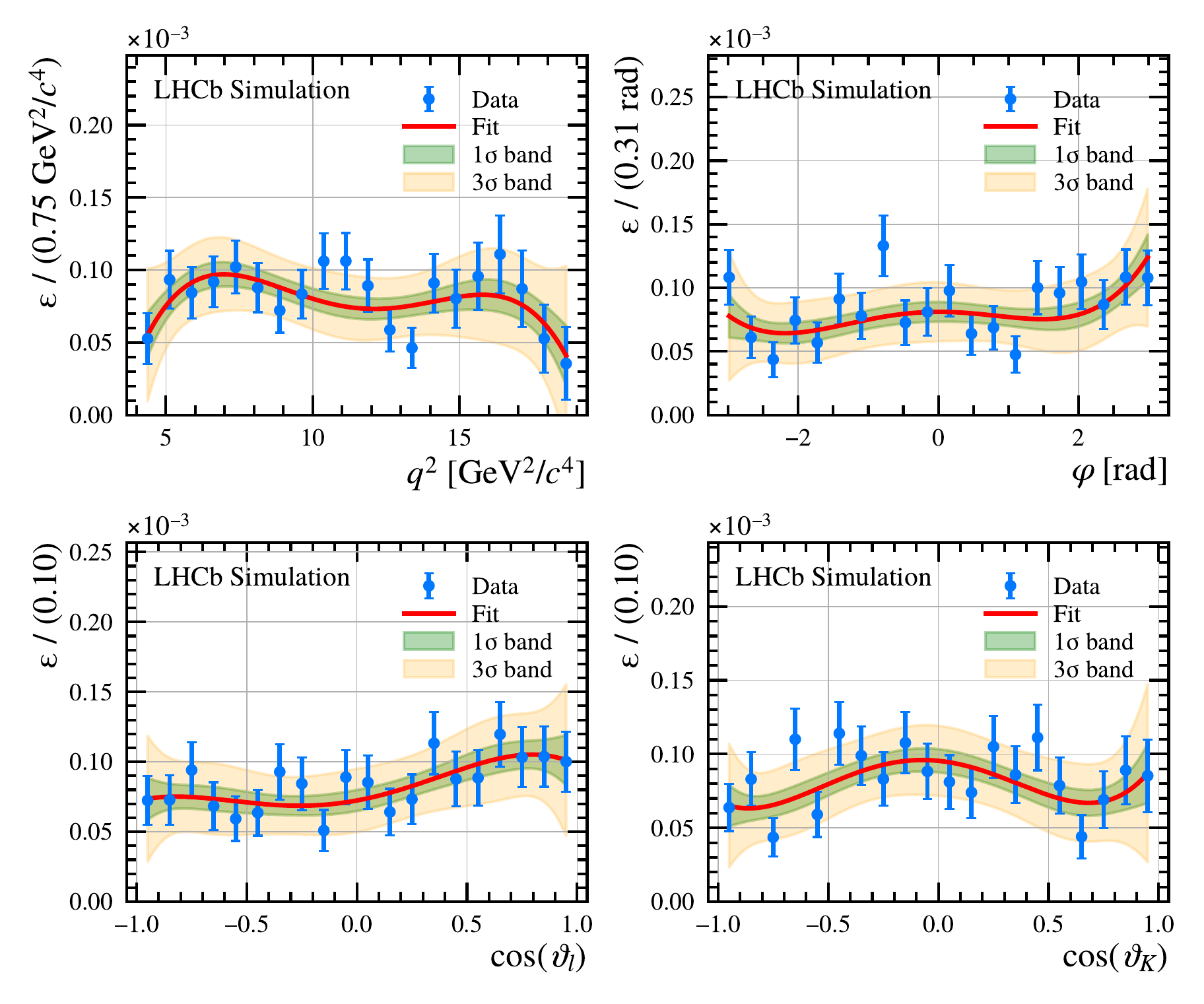}
    \caption{Efficiency distributions with their polynomial fit results as a function of (top left)~$q^2$, (top right)~$\varphi$, (bottom left)~$\cos(\theta_l)$ and (bottom right)~$\cos(\theta_K)$ for \SC and \tauToPiPiPiPiz decays.}
    \label{fig::np_efficiencies_sc_3pipi0}
\end{figure}

Efficiency as function of the  four-momentum transfer $q^2=m^2(\tau^{\pm} e^{\mp})$ and $m^2(K^{*0} \tau^{\pm} )$ (Fig.~\ref{fig:eff})  and as function of $q^2$ and the three decay angles $\varphi$, $\cos(\theta_l)$ and $\cos(\theta_K)$ (Figs.~\ref{fig::np_efficiencies_oc_3pi},~\ref{fig::np_efficiencies_oc_3pipi0},~\ref{fig::np_efficiencies_sc_3pi},~\ref{fig::np_efficiencies_sc_3pipi0}), defined as in Ref.~\cite{LHCb-PAPER-2013-019}, are presented. 

\FloatBarrier

\section{Beyond the Standard Model scenarios}\label{app:BSM} 
\begin{figure}[hbt!]
    \centering
    \includegraphics[width=0.48\linewidth]{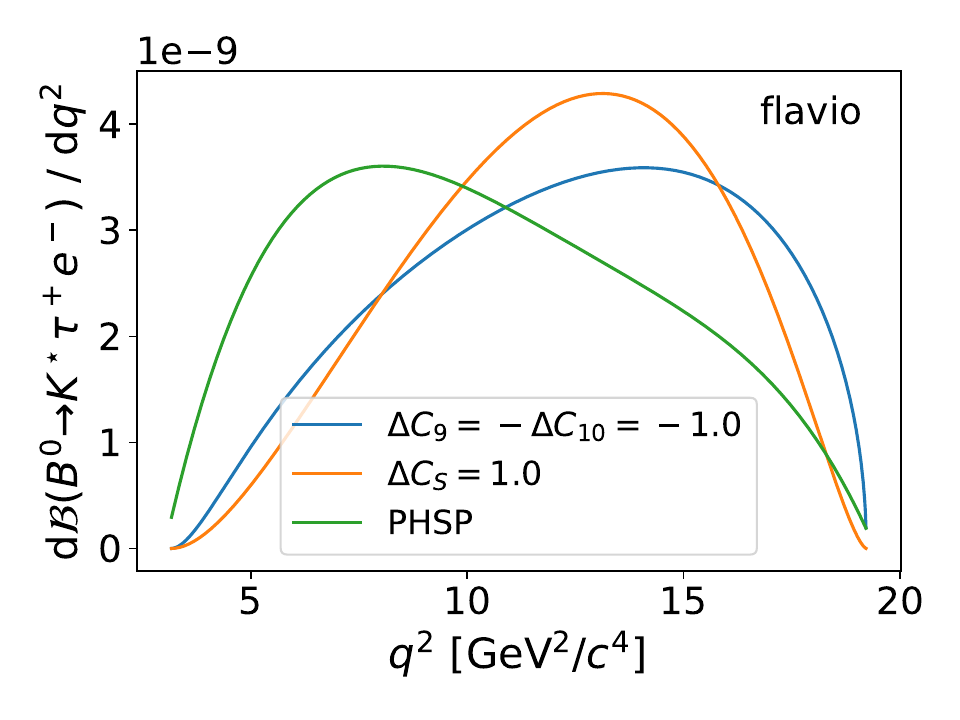}
    \includegraphics[width=0.48\linewidth]{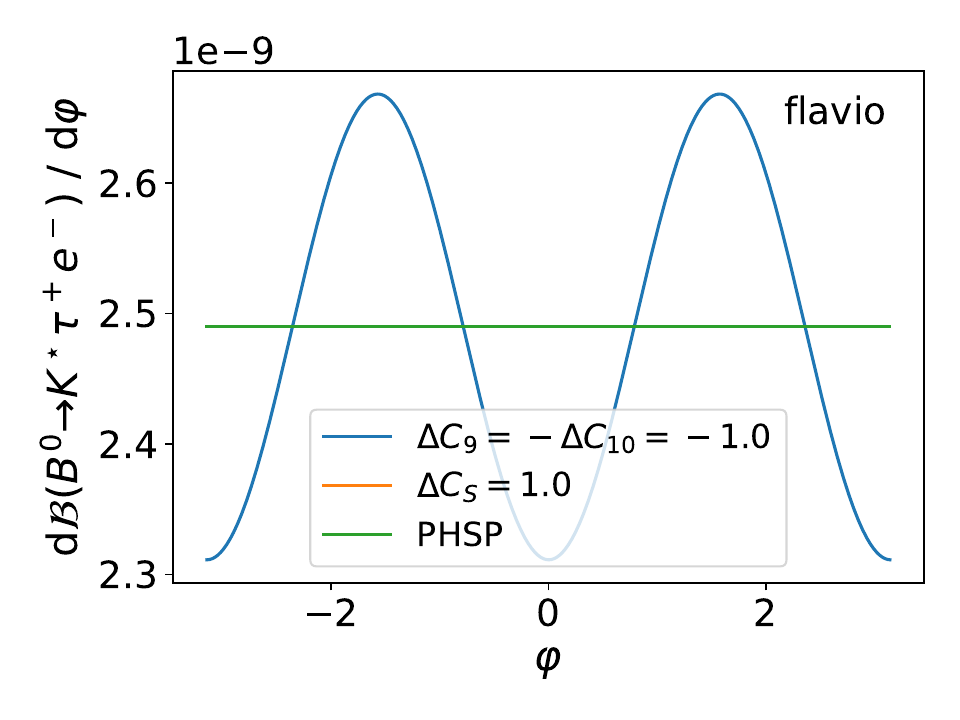}
    \includegraphics[width=0.48\linewidth]{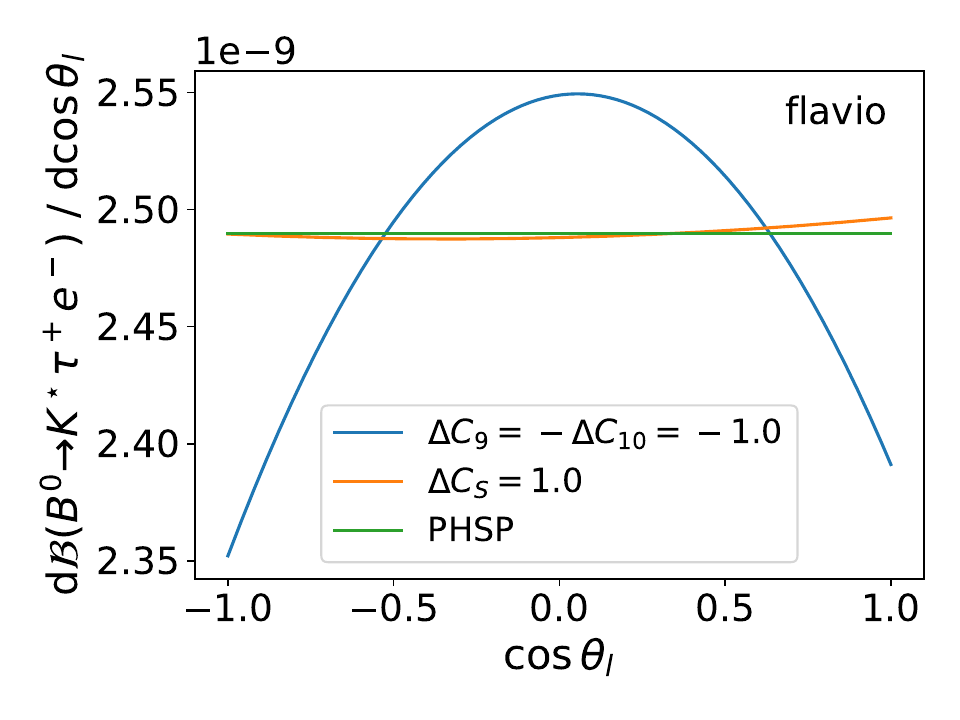}
    \includegraphics[width=0.48\linewidth]{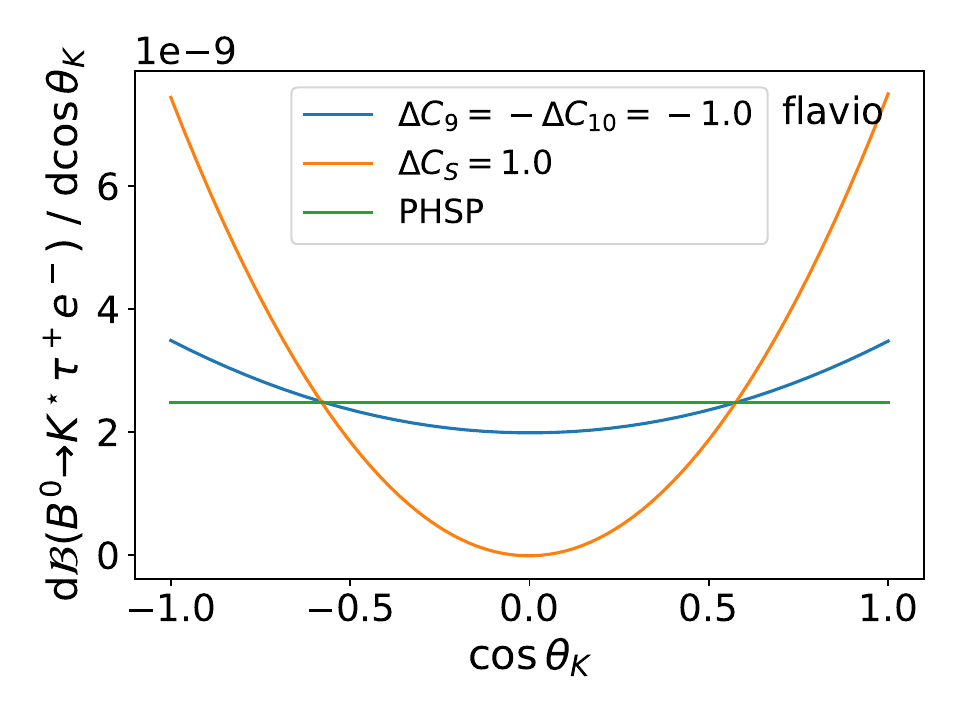}
    \caption{Differential decay rates as function of kinematics in different models for \OC decays.  
    Distributions are normalised to the integrated branching ratio obtained with the left-handed model ($\Delta C^{e\tau}_{9}=-\Delta C^{e\tau}_{10}=-1$).}
    \label{fig::np_models_oc}
\end{figure}

\begin{figure}[hbt!]
    \centering
        \includegraphics[width=0.48\linewidth]{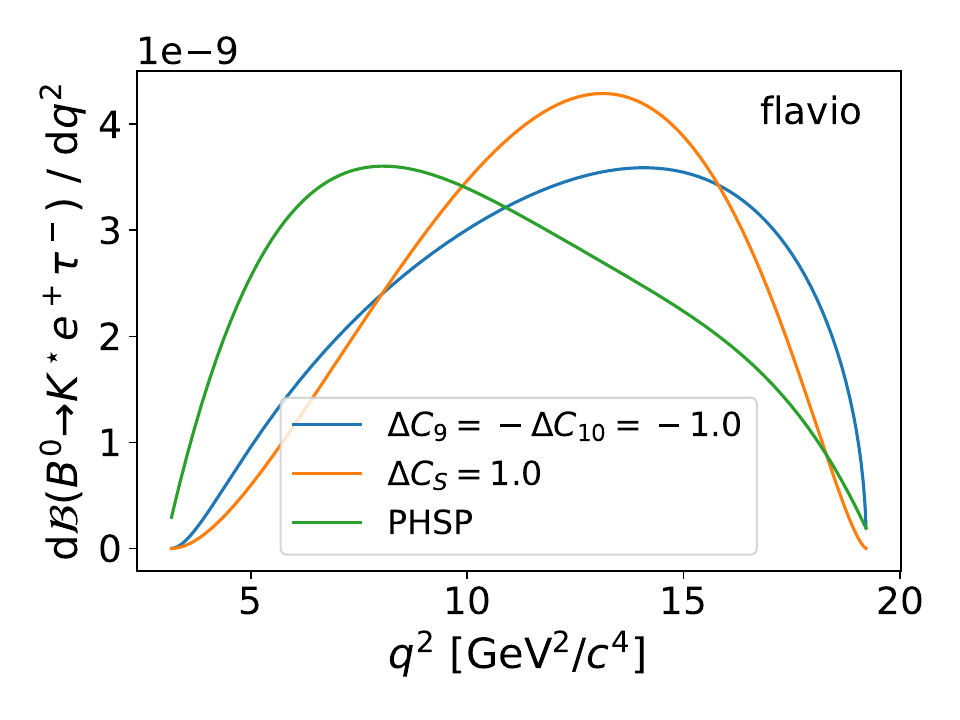}
    \includegraphics[width=0.48\linewidth]{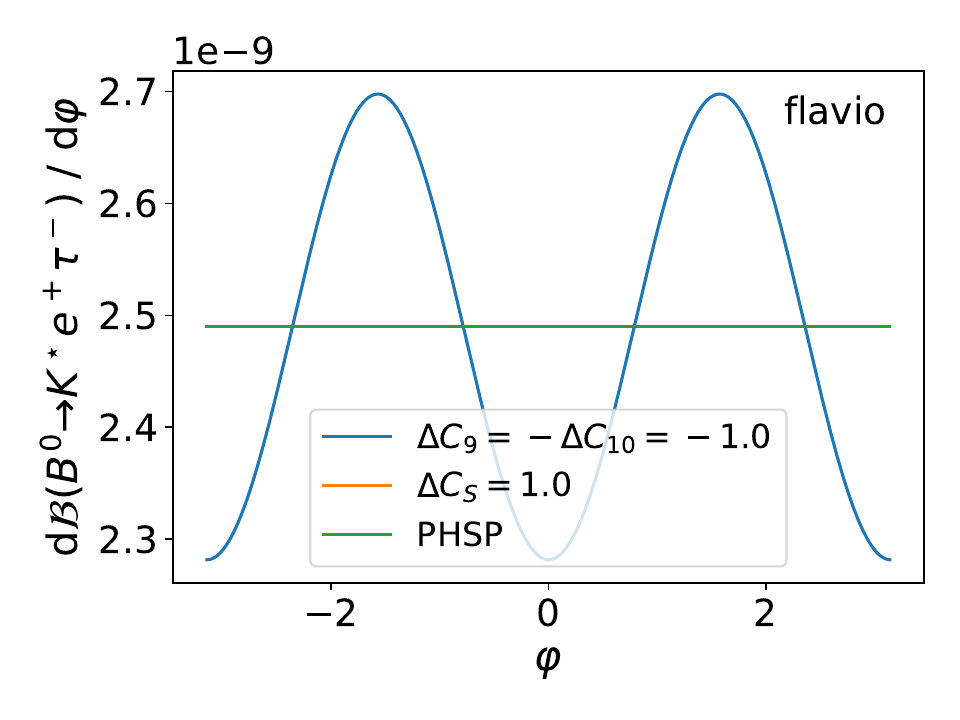}
    \includegraphics[width=0.48\linewidth]{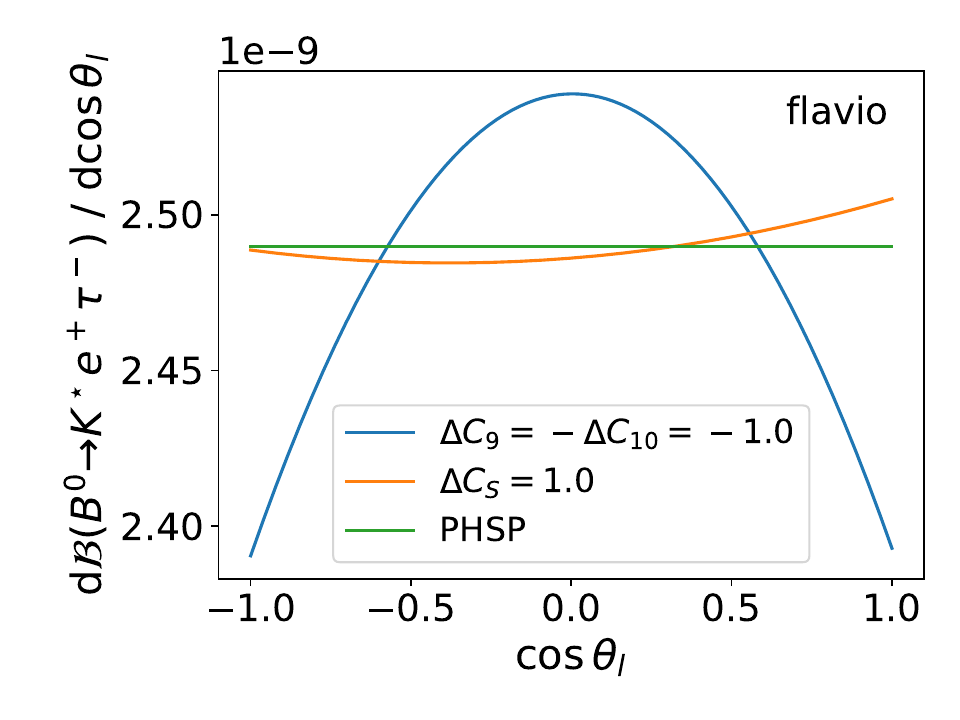}
    \includegraphics[width=0.48\linewidth]{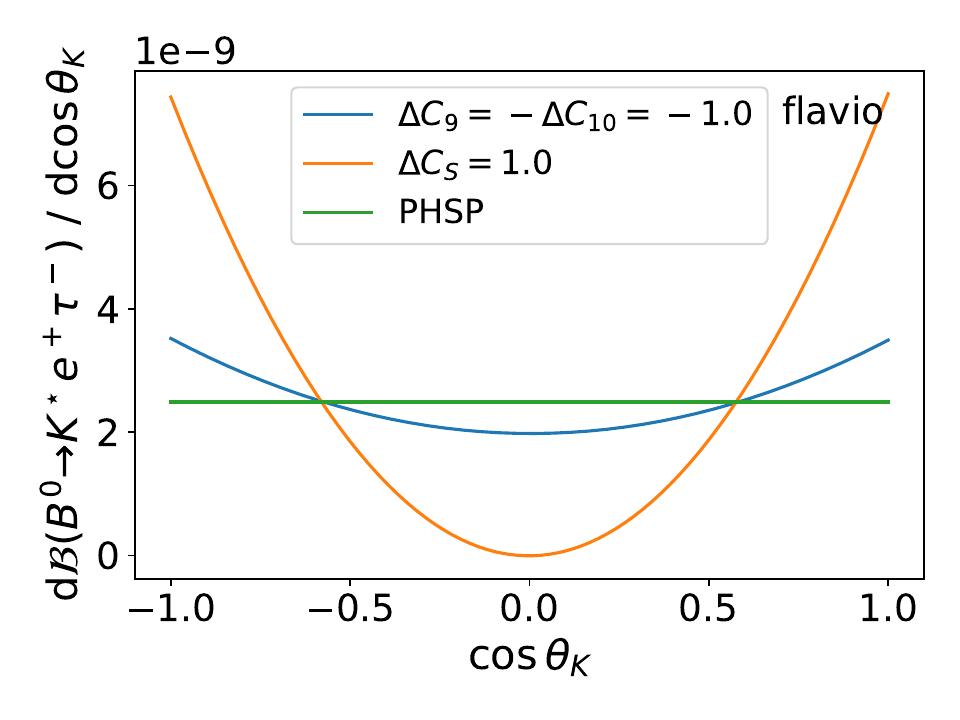}
    \caption{Differential decay rates as function of kinematics in different models for \SC decays. Distributions are normalised to the integrated branching ratio obtained with the left-handed model ($\Delta C^{e\tau}_{9}=-\Delta C^{e\tau}_{10}=-1$).  
    }
    \label{fig::np_models_sc}
\end{figure}

Scenarios beyond the Standard Model can result in very different distributions in four-momentum transfer $q^2=m^2(\tau^{\pm} e^{\mp})$ and the three decay angles $\varphi$, $\cos(\theta_l)$ and $\cos(\theta_K)$, defined as in Ref.~\cite{LHCb-PAPER-2013-019}. This is illustrated in Figs.~\ref{fig::np_models_oc} and \ref{fig::np_models_sc}, obtained using the python package \textsc{flavio}~\cite{straub2018flaviopythonpackageflavour}.

\clearpage


\addcontentsline{toc}{section}{References}
\bibliographystyle{LHCb}
\bibliography{main,standard,LHCb-PAPER,LHCb-CONF,LHCb-DP,LHCb-TDR}

\newpage
\centerline
{\large\bf LHCb collaboration}
\begin
{flushleft}
\small
R.~Aaij$^{38}$\lhcborcid{0000-0003-0533-1952},
A.S.W.~Abdelmotteleb$^{57}$\lhcborcid{0000-0001-7905-0542},
C.~Abellan~Beteta$^{51}$\lhcborcid{0009-0009-0869-6798},
F.~Abudin{\'e}n$^{57}$\lhcborcid{0000-0002-6737-3528},
T.~Ackernley$^{61}$\lhcborcid{0000-0002-5951-3498},
A. A. ~Adefisoye$^{69}$\lhcborcid{0000-0003-2448-1550},
B.~Adeva$^{47}$\lhcborcid{0000-0001-9756-3712},
M.~Adinolfi$^{55}$\lhcborcid{0000-0002-1326-1264},
P.~Adlarson$^{84}$\lhcborcid{0000-0001-6280-3851},
C.~Agapopoulou$^{14}$\lhcborcid{0000-0002-2368-0147}
% ...

{\footnotesize \it

% ...
$^{14}$Universit{\'e} Paris-Saclay, CNRS/IN2P3, IJCLab, Orsay, France\\
% ...
$^{38}$Nikhef National Institute for Subatomic Physics, Amsterdam, Netherlands\\
% ...
$^{47}$Instituto Galego de F{\'\i}sica de Altas Enerx{\'\i}as (IGFAE), Universidade de Santiago de Compostela, Santiago de Compostela, Spain\\
% ...
$^{51}$Physik-Institut, Universit{\"a}t Z{\"u}rich, Z{\"u}rich, Switzerland\\
% ...
$^{55}$H.H. Wills Physics Laboratory, University of Bristol, Bristol, United Kingdom\\
% ...
$^{57}$Department of Physics, University of Warwick, Coventry, United Kingdom\\
% ...
$^{60}$School of Physics and Astronomy, University of Glasgow, Glasgow, United Kingdom\\
$^{61}$Oliver Lodge Laboratory, University of Liverpool, Liverpool, United Kingdom\\
% ...
$^{69}$Syracuse University, Syracuse, NY, United States\\
% ...
$^{84}$Department of Physics and Astronomy, Uppsala University, Uppsala, Sweden, associated to $^{60}$\\
% ...
}
\end{flushleft}

\end{document}